\documentclass[12pt]{article} 

\usepackage{graphics,color}
\usepackage{amssymb}
\usepackage{amsbsy}
\usepackage{epsfig}

\newcommand{\be}{\begin{equation}}
\newcommand{\ee}{\end{equation}}
\newcommand{\bea}{\begin{eqnarray}}
\newcommand{\eea}{\end{eqnarray}}
\newcommand{\bit}{\begin{itemize}}
\newcommand{\eit}{\end{itemize}}
\newcommand{\bc}{\begin{center}}
\newcommand{\ec}{\end{center}}
\newcommand{\bra}{\langle}
\newcommand{\ket}{\rangle}
\newcommand{\zv}{{\mathbf 0}}
\newcommand{\si}{\,\mbox{$\sum$}\hs{-0.47cm}\int}
\newcommand{\sgn}{\mbox{sgn}}
\newcommand{\im}{{\mathrm{Im}}}
\newcommand{\re}{{\mathrm{Re}}}
\newcommand{\Tr}{{\mathrm{Tr}}}
\newcommand{\tr}{{\mathrm{tr}}}
\newcommand{\cO}{{\cal O}}

\newcommand{\cD}{{\cal D}}

\newcommand{\C}{{\cal C}}
\newcommand{\om}{\omega}

\newcommand{\tw}{\Gamma}

\newcommand{\ro}{r^{0}}
\newcommand{\lo}{l^{0}}
\newcommand{\po}{p^{0}}
\newcommand{\qo}{q^{0}}

\newcommand{\gv}{\boldsymbol\gamma}
\newcommand{\gz}{\gamma^{0}}

\newcommand{\lv}{{\mathbf l}}
\newcommand{\pv}{{\mathbf p}}
\newcommand{\kv}{{\mathbf k}}
\newcommand{\qv}{{\mathbf q}}
\newcommand{\rv}{{\mathbf r}}
\newcommand{\lvuni}{\hat{{\mathbf l}}}
\newcommand{\pvuni}{\hat{{\mathbf p}}}
\newcommand{\kvuni}{\hat{{\mathbf k}}}

\newcommand{\rvuni}{\hat{{\mathbf r}}}
\newcommand{\puni}{\hat{p}}

\newcommand{\hm}{\hspace*{-0.6cm}}
\newcommand{\hs}[1]{\hspace*{#1}}

\newcommand{\half}{\frac{1}{2}}

\newcommand{\ep}{\epsilon}

\newcommand{\bean}{\begin{eqnarray*}}
\newcommand{\eean}{\end{eqnarray*}}
\newcommand{\nn}{\nonumber}

\newcommand{\vecnul}{{\mathbf 0}}

\newcommand{\Dslash}{/\!\!\!\!D}

\newcommand{\Lslash}{/\!\!\!\!L}
\newcommand{\Pslash}{/\!\!\!\!P}

\newcommand{\Rslash}{/\!\!\!\!R}

\newcommand{\pbp}{{\bar p_+}}
\newcommand{\pbm}{{\bar p_-}}
\newcommand{\li}{{\rm Li}}

\begin{document}

\title{
\vskip -100pt
{
\begin{normalsize}
\mbox{} \hfill SWAT 05-427\\
\mbox{} \hfill hep-ph/0503161\\
\vskip  30pt
\end{normalsize}
}
{\bf\Large
Transport coefficients in large $N_{f}$ gauge theories\\ 
with massive fermions
}
\author{
\addtocounter{footnote}{2}
Gert Aarts$^{\natural,}$\footnote{current address, email: g.aarts@swan.ac.uk}
\addtocounter{footnote}{2}
 {} and
Jose M.\ Mart{\'\i}nez Resco$^{\natural,}$\thanks{
current address, email: martinezrescoj@brandonu.ca}
 \\ {} \\
 {}$^\natural${\em\normalsize Department of Physics, The Ohio State 
University} \\
 {\em\normalsize 174 West 18th Avenue, Columbus, OH 43210, USA}
 \\ {} \\
 {}$^\ddagger${\em\normalsize Department of Physics, University of Wales 
Swansea} \\
 {\em\normalsize Singleton Park, Swansea, SA2 8PP, United Kingdom}
 \\ {} \\
 {}$^\|${\em\normalsize Department of Physics \& Astronomy, 
Brandon University} 
 \\
 {\em\normalsize Brandon, Manitoba R7A 6A9, Canada }
}
}
\date{March 16, 2005}
\maketitle

\begin{abstract}
We compute the shear viscosity and the electrical conductivity in gauge 
theories with massive fermions at leading order in the large $N_f$ 
expansion. The calculation is organized using the $1/N_f$ expansion of the 
2PI effective action to next-to-leading order. We show explicitly that the 
calculation is gauge fixing independent and consistent with the Ward 
identity. We find that these transport coefficients depend in a 
nontrivial manner on the coupling constant and fermion mass. For large 
mass, both the shear viscosity and the electrical conductivity go to zero.

\end{abstract}

\newpage


\section{Introduction}

Transport coefficients in relativistic gauge theories have been discussed 
in a number of papers in the past few years 
\cite{Arnold:2000dr,Policastro:2001yc,Moore:2001fg,ValleBasagoiti:2002ir,Aarts:2002tn,Boyanovsky:2002te,Buchel:2004di,Defu:2005hb,Peshier:2005pp}.
The motivation comes from possible applications in heavy ion physics 
and the early universe, as well as from theoretical interest. However, the 
attention has mostly been focused on ultrarelativistic theories, where the 
scale is set exclusively by the temperature. In this paper we undertake 
the computation of transport coefficients in gauge theories at 
temperatures where the fermion mass cannot be neglected. We carry out this 
study in the large $N_f$ limit of QED and QCD, where a complete leading 
order calculation is possible. For massless fermions transport 
coefficients have been computed in large $N_f$ gauge theories in 
Ref.~\cite{Moore:2001fg}, using kinetic theory. A study of thermodynamic 
properties of gauge theories in the large $N_f$ limit can be found in 
Ref.~\cite{Moore:2002md}.

We perform a diagrammatic calculation, organized using the $1/N_f$ 
expansion of the two-particle irreducible (2PI) effective action to 
next-to-leading order (NLO). The 2PI effective action is a useful tool in 
studying the nonequilibrium dynamics of quantum fields 
\cite{Berges:2000ur}. In actual applications, the 2PI effective action is 
truncated at some order in a chosen expansion parameter. In 
Ref.~\cite{Aarts:2003bk} it was shown for a number of theories that the 
lowest nontrivial truncations correctly determine transport coefficients 
at leading (logarithmic) order in the expansion parameter. Here we show 
explicitly that the lowest order nontrivial truncation of the 2PI 
effective action in the $1/N_f$ expansion provides all the required 
ingredients to successfully compute the shear viscosity and the electrical 
conductivity. When considering gauge theories and effective actions, care 
is required with respect to gauge invariance and Ward identities 
\cite{Arrizabalaga:2002hn}. We show that despite the nontrivial 
resummation of diagrams carried out, the method provides a gauge fixing 
independent result and is consistent with the Ward identity. This provides 
an explicit example of a nontrivial quantity for which potential non gauge 
invariant contributions in a fully self-consistent calculation would be 
suppressed by powers of the expansion parameter.

\begin{figure}[t]
 \centerline{\epsfig{figure=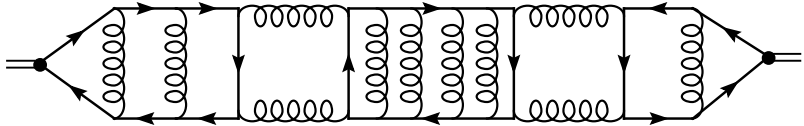,height=1.6cm}}
 \caption{Typical skeleton ladder diagram that contributes to the shear 
viscosity in large $N_f$ QCD at leading order in the $1/N_f$ expansion.}
 \label{figladder}
\end{figure}

The paper is organized as follows. In Section~\ref{2pi}, we formulate the 
2PI effective action to NLO in large $N_f$ QED. We obtain the integral 
equation relevant for the calculation of the shear viscosity and the 
electrical conductivity and discuss powercounting in the $1/N_f$ 
expansion. We show that a typical diagram that contributes to the shear 
viscosity at leading order in the large $N_f$ expansion is as shown in 
Fig.~\ref{figladder}. Plasma effects relevant for transport coefficients 
are studied in Section~\ref{quasiparticles}. In the following Section, we 
explicitly work out the integral equation relevant for the shear viscosity 
and write it in a form convenient for a variational treatment. In 
Section~\ref{conductivity}, this analysis is repeated for the electrical 
conductivity and the Ward identity is explicitly checked. We generalize 
the discussion from QED to large $N_f$ QCD in Section \ref{color}. The 
numerical analysis and our results are presented in 
Section~\ref{solution}. The final Section is devoted to the conclusions. 
In Appendix~\ref{appendixA} we derive a set of integral equations from the 
2PI effective action which are employed in the main text. Finally, 
Appendix~\ref{appendixB} contains parametric estimates in the leading 
logarithmic approximation, for both massless and very heavy fermions.

A short summary of these results has appeared in Ref.~\cite{Aarts:2004xs}. 
Part of the analysis is very similar to the study of the shear viscosity 
in the $O(N)$ model in the large $N$ limit \cite{Aarts:2004sd}. When 
possible, we will refer to that paper for further details.


\section{2PI-$1/N$ expansion} 
\label{2pi}

Since the structure of QED and QCD is similar in the large $N_{f}$ limit, 
we use QED in the following for the purpose of discussion. Color 
factors will be introduced later. The action for $N_f$ identical fermion 
fields $\psi_a$ ($a=1,\ldots,N_f$) then reads\footnote{We use 
$g_{\mu\nu}=\mbox{diag}(+,-,-,-)$, so that $P^2=p_0^2-p^2$, $p=|\pv|$. The 
$\gamma$-matrices obey $\{\gamma^\mu, \gamma^\nu\}=2g^{\mu\nu}$. Traces 
over Dirac indices are indicated with $\tr$.}
\be
 S = \int_x \left[ -\frac{1}{4}F_{\mu\nu}F^{\mu\nu} 
 +\bar\psi_a \left(i\Dslash - m \right)\psi_a \right] 
 + S_{\rm gf} +S_{\rm gh},
\ee
with 
\be
 \label{eqcov}
 \Dslash = \gamma^\mu D_\mu, \;\;\;\;\;\;\;\; 
 D_\mu = \partial_\mu + \frac{ie}{\sqrt{N_f}} A_\mu,
\ee
and we use the notation
\be
 \int_x = \int_\C dx^0 \int d^3 x,
\ee
where $\C$ is a contour in the complex-time plane. Note that we have 
rescaled the coupling constant with $\sqrt{N_f}$, so that in the large 
$N_f$ limit $N_f$ goes to infinity while $e$ remains finite (after renormalization). 
To fix the gauge we use a general linear gauge fixing 
condition,
\be
 S_{\rm gf} = -\int_x \frac{1}{2\xi} (f_\mu A^\mu)^2.
\ee
Below we specialize to the generalized Coulomb gauge: $f_0=0, 
f_i=\partial_i$. The ghost part is not needed explicitly.

The 2PI effective action is an effective action for the contour-ordered 
two-point functions
\be
 D_{\mu\nu}(x,y) = \bra T_\C A_\mu(x) A_\nu(y) \ket, 
 \;\;\;\;\;\;\;\;
 S_{ab}(x,y) = \bra T_\C \psi_a(x)\bar\psi_b(y)\ket,
\ee
and can be written as \cite{Cornwall:1974vz}
\bea
 \nn
 \Gamma[S, D] = &&\hm 
 \frac{i}{2}\Tr \ln D^{-1} + \frac{i}{2}\Tr\, D_0^{-1}(D-D_0) 
 \\&&\hm
 -i\Tr \ln S^{-1} -i\Tr\, S_0^{-1}(S-S_0)  
 +\Gamma_2[S,D] + \mbox{ghosts},
\eea
 where $D_0^{-1}$ and $S_0^{-1}$ are the free inverse propagators. The 2PI 
effective action framework automatically entails the existence of a set of 
coupled integral equations for the various 4-point functions. These 
integral equations contain the relevant physics for the calculation of 
some transport coefficients in a number of theories~\cite{Aarts:2003bk}. 
In Appendix \ref{appendixA} we briefly describe how to obtain the relevant 
set in the theory we study here.

\begin{figure}[t]
 \centerline{\epsfig{figure=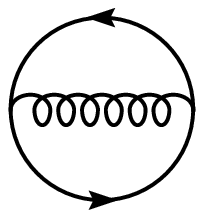,height=1.6cm}}
 \caption{NLO contribution to the 2PI effective action in 
          the $1/N_f$ expansion.}
 \label{fig:2pi-lo}
\end{figure}

\begin{figure}[t]
 \centerline{\epsfig{figure=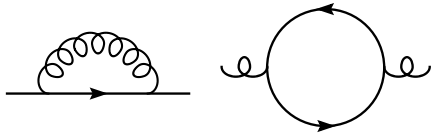,height=1.6cm}}
 \caption{Fermion and gauge boson self energy.}
 \label{fig:selfenergy}
\end{figure}

The lowest order contribution to $\Gamma_2[S,D]$ appears at 
next-to-leading order (NLO) in the large $N_f$ expansion (see 
Fig.~\ref{fig:2pi-lo})
\be
 \Gamma_2^{\rm NLO}[S,D] = -\frac{ie^2}{2N_f}\int_{xy} \tr\,
 \gamma^\mu S_{ab}(x,y)\gamma^\nu S_{ba}(y,x)D_{\mu\nu}(x,y).
\ee
We specialize to the completely symmetric case and write
$S_{ab}=\delta_{ab}S$, $\Sigma_{ab}=\delta_{ab}\Sigma$. The resulting self 
energies (see Fig.~\ref{fig:selfenergy}) are then
\bea 
 \label{eqpi0}
 \Pi^{\mu\nu}(x,y)=&&\hm 
   e^2\tr\, \gamma^\mu S(x,y)\gamma^\nu S(y,x),
 \\  
 \label{eqsigma0}
 \Sigma(x,y)=&&\hm -\frac{e^2}{N_f} \gamma^\mu S(x,y)\gamma^\nu 
D_{\mu\nu}(x,y).
\eea
They depend on the full propagators, determined by the Dyson equations
\be
 D^{-1} = D_0^{-1} -\Pi, \hs{3cm} S^{-1} = S_0^{-1} -\Sigma.
\ee

\begin{figure}[t]
 \centerline{\epsfig{figure=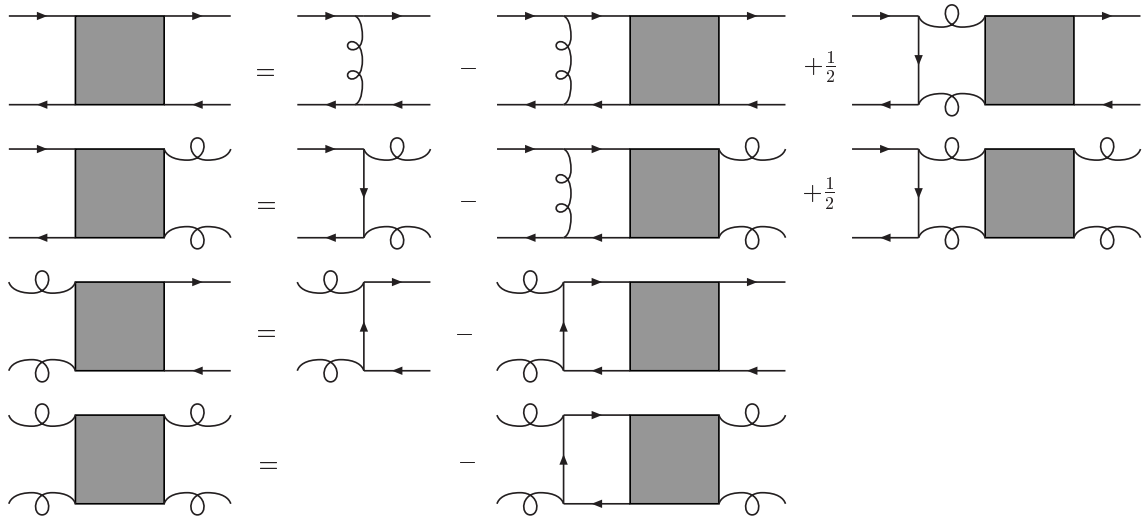,height=7.0cm}}
 \caption{Integral equations for the 4-point functions at NLO in the 
          $1/N_f$ expansion.}
 \label{fig:ie}
\end{figure}

The set of integral equations for the 4-point functions (see 
Eq.~(\ref{eqfourex}) in Appendix~\ref{appendixA}), up to this order in the 
large $N_f$ expansion, is shown in Fig.~\ref{fig:ie}. These coupled 
equations sum all the diagrams that are required to obtain the shear 
viscosity and electrical conductivity at leading order in the $1/N_{f}$ 
expansion \cite{Aarts:2003bk}. This can be argued as follows. Kubo 
formulas relate these transport coefficients to the slope of 
current-current spectral functions at vanishing frequency
\be
 \label{eta}
 \eta = \frac{1}{20}\frac{\partial}{\partial q^0}
 \rho_{\pi\pi}(q^0,\vecnul)\Big|_{q^0=0}, 
 \hs{1cm} 
 \sigma = \frac{1}{6}\frac{\partial}{\partial \qo}
 \rho_{jj}(\qo,\vecnul)\Big|_{\qo=0},
\ee 
where the spectral functions are defined as
\be 
 \rho_{\pi\pi}(x-y)=\bra [\pi_{ij}(x), \pi_{ij}(y)]\ket, \hs{1cm}
 \rho_{jj}(x-y)=\bra[j^{i}(x),j^{i}(y)]\ket.
\ee 
Here $\pi_{ij}$ is the traceless part of the spatial energy-momentum 
tensor,
\be \label{pi} 
 \pi_{ij}(x)=F_{i\mu}F_{j\mu}-\frac{1}{3}\delta_{ij}F_{k\mu}F_{k\mu}-
 i\bar{\psi}_a \left( \frac{\gamma_{i}D_{j}-\gamma_{j}D_{i}}{2}
 -\frac{1}{3}\delta_{ij}\gamma_{k}D_{k} \right) \psi_a, 
\ee 
and $j^i(x)=q_f\bar\psi_a(x)\gamma^i\psi_a(x)$ is the electromagnetic 
current, with $q_f$ the charge of the fermion.\footnote{This is the charge 
with which the fermions couple to the external operator; we prefer to 
distinguish it from the coupling between the gauge bosons and the fermions 
in the ladder diagrams. In QED it is also rescaled, so that $q_f =e/\sqrt{N_f}$, 
while in QCD it is not.} 
The correlators in Kubo formulas are computed in thermal equilibrium, 
so from now on we specialize to the Matsubara contour and work in 
momentum space. 

\begin{figure}[t]
 \centerline{\epsfig{figure=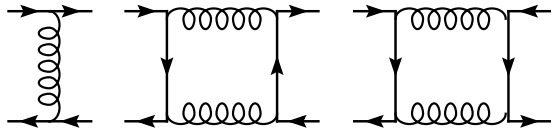,height=1.6cm}}
 \caption{Rungs in the integral equation for the fermion 4-point function.
}
 \label{fig:rungs}
\end{figure}

The correlators in the Kubo formulas are required in a specific kinematic 
configuration which, as is well known, causes them to suffer from 
so-called pinching poles. These pinching poles are screened by the 
imaginary part of self energy, leading to the appearance of a factor 
inversely proportional to this imaginary part. This modifies the naive 
power counting scheme. The fermionic one loop diagram, which contributes 
to both transport coefficients, is naively of order $N_f$, due to the 
$N_f$ identical fermion fields that run in the loop. Because of the 
pinching poles, this is enhanced by the inverse thermal width (given by 
the imaginary part of self energy) which is of order $1/N_f$, as we show 
below. We find therefore that the conductivity and the shear viscosity are 
proportional to $N_{f}^{2}$ in the large $N_f$ limit (apart from the 
external charges in the case of the electrical conductivity). Adding a 
vertical photon line to the one-loop fermion diagram gives a contribution 
that is of the same order; the factor of $1/N_f$ from the added vertices 
is compensated by a new pair of propagators with pinching poles and a new 
inverse factor of the thermal width. This remains true when adding any 
number of vertical photon lines; therefore all these diagrams have to be 
summed. A contribution of the same order is also obtained when considering 
a box rung with horizontal photon lines and vertical fermion lines (see 
Fig.\ \ref{fig:rungs}). In this case, a new pair of propagators with 
pinching poles along with a new closed fermion loop compensates for the 
additional four coupling vertices. Note that there are two ways a box rung 
can be added, depending on the orientation of the fermion lines. Again, a 
diagram with any number of box rungs contributes at leading order as well. 
These kind of diagrams are precisely those which are summed by the 
integral equations for the fermionic 4-point function we obtained from the 
2PI effective action.

In the case of the shear viscosity, the external operator also couples to 
two gauge boson fields. It is therefore necessary to consider diagrams 
with gauge bosons on the side rails. Again, the corresponding imaginary 
part of the gauge boson self energy has to be included in the side rails 
propagators to avoid pinching 
poles. If the gauge boson is an onshell stable excitation, its self energy 
in Fig.\ \ref{fig:selfenergy} yields a thermal width only when at least 
one of the fermion lines in the diagram is dressed, a contribution of 
order $1/N_f$ (see Eq.\ (\ref{eqsigma0})). As a result the gauge boson 
thermal width is of order $1/N_f$, similar as the fermion thermal width. 
On the other hand, if the gauge boson is an unstable excitation, no 
fermion lines need to be dressed in the gauge boson self energy to get a 
non-vanishing imaginary part, which is therefore of order $N_f^0$; in this 
case pinching poles do not lead to a further enhancement. 
However, in both cases there is only one gauge boson compared to $N_f$ 
fermion fields. Therefore these diagrams are subleading in the $1/N_f$ 
expansion. For the shear viscosity we finally also have to consider 
diagrams where one external operator couples to two gauge boson fields and 
the other one to two fermion fields, and there is at least one fermion 
rung. In this case there are pinching poles from the pair of gauge boson 
propagators and from the pair of fermion propagators. If the gauge boson 
is an onshell stable excitation, we find two powers of $N_f$ from the 
pinching poles, one power of $N_f$ from the closed fermion loop and a 
power of $1/N_f$ from the coupling vertices. However, due to kinematics 
this diagram gives a nonzero contribution only when the spectral density 
of the fermionic rung is offshell, which introduces a further power of 
$1/N_f$ and makes the contribution from this diagram subleading. If the 
gauge boson is an unstable excitation, the fermionic rung can be onshell 
but we find only one power of $N_f$ from the pinching poles and the 
diagram is subleading as well. We conclude that diagrams where one or both 
of the external operators couple to gauge bosons can be neglected at 
leading order in the large $N_f$ expansion.

It is therefore not necessary to consider the full set of integral 
equations in Fig.~\ref{fig:ie}; only the closed integral equation for the 
4-point function where all external legs are fermionic is required. In 
this respect, the large $N_f$ computation is slightly easier than the 
leading-log calculation in the weak-coupling limit, where two coupled 
integral equations for the fermion and the gauge boson contributions have 
to be solved simultaneously \cite{ValleBasagoiti:2002ir}. Instead it is 
very similar to the analysis in the $O(N)$ model in the large $N$ limit, 
with the gauge boson and the bubble chain playing a similar 
role~\cite{Aarts:2004sd}.

The individual kernels in the integral equations in Fig.\ \ref{fig:ie} are 
obtained by cutting one line in the self energies and read 
\bea
\nn \Lambda_{ab;cd}(R,P) =&&\hm -\frac{e^2}{N_f} \delta_{ad}\delta_{bc} 
 \gamma^\mu D_{\mu\nu}(R-P)\gamma^\nu, \\
 \Lambda_{ab;\mu\nu}(R,P)=&&\hm \frac{e^2}{N_f} \delta_{ab}\left[ \gamma^\mu 
 S(R-P)\gamma^\nu+\gamma^\nu S(R-P)\gamma^\mu \right], \\
\nn \Lambda_{\mu\nu;ab}(R,P)=&&\hm \frac{e^2}{N_f} \delta_{ab} 
 \left[ \gamma^\mu S(P-R)\gamma^\nu+\gamma^\nu S(P-R)\gamma^\mu \right],
\eea
where $R$ is the momentum that enters and leaves on the left and $P$ 
enters and leaves on the right. To obtain a closed integral equation for 
the fermionic 4-point function, the third equation in Fig.~\ref{fig:ie} is 
substituted into the first one, leading to 
\bea
 \Gamma^{(4)}_{ab;cd}(R,K) = \widetilde{\Lambda}_{ab;cd}(R,K) - \si_P 
 \widetilde{\Lambda}_{ab;a'b'}(R,P)S(P)S(P)\Gamma^{(4)}_{b'a';cd}(P,K),
\eea
with the effective kernel
\bea
 &&\hm \widetilde{\Lambda}_{ab;cd}(R,P) = 
 \Lambda_{ab;cd}(R,P)+\half \si_L \Lambda_{ab;\mu\nu}(R,L) 
 D_{\nu\sigma}(L) D_{\rho\mu}(L) \Lambda_{\rho\sigma;cd}(L,P) 
 \nn\\ 
  &&\hm = -\frac{e^2}{N_f} 
 \delta_{ad}\delta_{bc} \gamma^\mu D_{\mu\nu}(R-P)\gamma^\nu 
 \\ &&\hm  \nn
   + \frac{e^4}{N_f^2} \delta_{ab}\delta_{cd}\si_L 
 \left[ \gamma^\nu S(R-L) \gamma^\mu +\gamma^\mu S(R+L) \gamma^\nu 
 \right]D_{\nu\sigma}(L) D_{\rho\mu}(L)\gamma^\rho S(P-L) \gamma^\sigma.
\eea
We use the notation
\be
 \si_P=T\sum_n\int_\pv, \hs{3cm} \int_\pv=\int\frac{d^3p}{(2\pi)^3},
\ee
where the sum runs over the corresponding Matsubara frequencies. 
To carry out the frequency sums, it is convenient to introduce a 3-point 
effective vertex $\Gamma_{ab}(P+Q,P)$ as
\begin{figure}[t]
 \centerline{\epsfig{figure=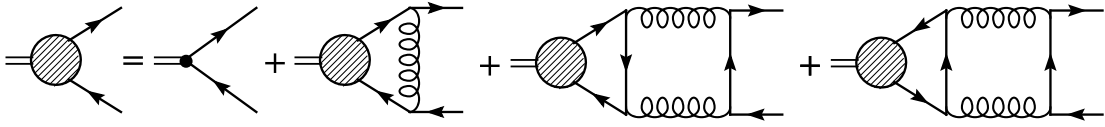,height=1.55cm}}
 \caption{Integral equation for the full 3-point function.}
 \label{fig:ie-f}
\end{figure}
\be
 \Gamma_{ab}(P+Q,P) = \Gamma_{ab}^0(\pv) - 
  \si_R S(R+Q)\Gamma_{cd}^0(\rv)S(R)\Gamma^{(4)}_{cd;ab}(R,P;Q),
\ee
 where $\Gamma_{ab}^0(\pv)$ is the bare coupling between the fermion 
fields and the external operator under consideration, and $Q$ is the 
momentum entering the operator insertion. This yields the final integral 
equation we work on in the remainder of this paper (see 
Fig.~\ref{fig:ie-f}) 
\be \label{ie}
 \Gamma_{ab}(P+Q,P) = \Gamma_{ab}^0(\pv) - \si_R S(R+Q)\Gamma_{cd}(R+Q,R)S(R) 
 \widetilde{\Lambda}_{cd;ab}(R,P;Q),
\ee
with the kernel
\bea  \label{kernel}
 &&\hm\widetilde{\Lambda}_{ab;cd}(R,P;Q) = -\frac{e^2}{N_f} \delta_{ad} 
 \delta_{bc} \gamma^\nu D_{\mu\nu}(R-P)\gamma^\mu 
 + \frac{e^4}{N_f^2} \delta_{ab}\delta_{cd}
 \\  \nn
 &&\hm \times\si_L \left[ \gamma^\nu S(R-L)\gamma^\mu  
 + \gamma^\mu S(R+L+Q) \gamma^\nu \right] 
 D_{\mu\rho}(L+Q) D_{\sigma\nu}(L) \gamma^\rho S(P-L)\gamma^\sigma,
\eea
A typical skeleton diagram that contributes to the shear viscosity in 
large $N_f$ gauge theories is depicted in Fig.~\ref{figladder}. Throughout 
the paper we drop subleading powers of~$N_f$.


\section{Quasiparticles}
\label{quasiparticles}

In this section we study the effects of the medium on the propagation of 
both fermions and gauge bosons at this order in the $1/N_f$ expansion. In 
particular, we discuss the fermionic thermal width and compute the full 
gauge boson self energy required at this order in the $N_f$ limit.


\subsection{Gauge boson}

We choose to work with the gauge boson propagator in the generalized Coulomb gauge, 
so that it reads
\be
 D^{\mu\nu}(P) = P_T^{\mu\nu}\Delta_T(P) + g^{\mu0}g^{\nu0}\Delta_L(P) + 
 \xi\frac{P^\mu P^\nu}{p^4}, 
\ee
with transverse and longitudinal propagators
\bea
 \Delta_T(P) = &&\hm \frac{1}{\om_n^2+p^2+\Pi_T(P)} = 
 -\int_{-\infty}^{\infty} \frac{d\om}{2\pi} 
 \frac{\rho_T(\om,\pv)}{i\om_n-\om}, \\ 
 \Delta_L(P) = &&\hm \frac{-1}{p^2+\Pi_L(P)} = -\frac{1}{p^2} - 
 \int_{-\infty}^{\infty} \frac{d\om}{2\pi}
 \frac{\rho_L(\om,\pv)}{i\om_n-\om}.
\eea
In this gauge the gauge boson spectral function is independent of $\xi$, 
\be
 \rho^{\mu\nu}(P) = P_T^{\mu\nu}\rho_T(P) + g^{\mu 0} g^{\nu 0}\rho_L(P).
\ee
The self energy 
\be \label{se-pi}
 \Pi^{\mu\nu}(P) = e^2 
 \si_K \tr\, \gamma^\mu S(P+K)\gamma^\nu S(K),    
\ee
is decomposed as
\be
 \Pi^{\mu\nu}(P) = P_T^{\mu\nu}\Pi_T(P)+\frac{P^2}{p^2} P_L^{\mu\nu}\Pi_L(P),
\ee
with the usual projectors
\be
 P_T^{ij} = \delta^{ij}-\puni^i\puni^j, \;\;\;\;\;\;\;\;
 P_T^{\mu 0} = P_T^{0\nu} = 0, \;\;\;\;\;\;\;\;
 P_L^{\mu\nu} = g^{\mu\nu} -\frac{P^\mu P^\nu}{P^2} + P_T^{\mu\nu}.
\ee
The transverse and longitudinal self energies are then
\be
 \Pi_{L} = -\Pi^{00}, \hs{1.5cm} 
 \Pi_{T} = -\half\left(\Pi^{\mu}_{\mu}+\frac{P^{2}}{p^{2}}\Pi^{00}\right).
\ee
Since we drop subleading corrections in the $1/N_f$ expansion, and 
pinching poles are not an issue here, the fermionic propagators in 
Eq.~(\ref{se-pi}) can be taken at leading order, i.e. free ones.

We need to compute both $\Pi^{00}$ and $\Pi^{\mu}_{\mu}$. 
We split the self energy into vacuum and thermal parts
\be
 \Pi^{\mu\nu}=\Pi^{\mu\nu}_{\rm vac}+\Pi^{\mu\nu}_{\rm th},
\ee
where the vacuum contribution has the usual form
\be
 \Pi^{\mu\nu}_{\rm vac}(P) = \left(P^2 g^{\mu\nu}-P^\mu 
 P^\nu\right)\Pi_{\rm vac}(P), 
\ee
with
\be
 \Pi_{\rm vac}(P)=\frac{e_0^2\mu^{-2\ep}}{12\pi^2} 
 \left(\frac{1}{\ep}+\ln4\pi-\gamma_E\right)+\Pi^f_{\rm vac}(P).
\ee
We used dimensional regularization in $3-2\epsilon$ dimensions and 
$e_{0}$ is the bare coupling constant. In order to carry out the 
renormalization,\footnote{For renormalization of 2PI effective 
actions beyond what is needed here, see Ref.\ 
\cite{vanHees:2001ik}.}
we introduce the dimensionless running coupling 
constant in the $\overline{MS}$ scheme 
\be
 \frac{1}{e^2(\mu)} \equiv \frac{\mu^{2\ep}}{e_0^2} + 
 \frac{1}{12\pi^2}\left(\frac{1}{\ep} + \ln 4\pi - \gamma_E \right).
\ee
The running coupling constant obeys the usual renormalization group 
equation with $\beta(e^{2})=e^{4}/(6\pi^{2})$. Renormalization is now 
straightforward,
\be
 \mu^{-2\ep}e_0^2\Delta_{T/L}^{\rm bare} = e^2(\mu)\Delta_{T/L},
\ee
with the renormalized propagators,
\be  \label{finiteprop}
 \Delta_T = \frac{1}{-P^2(1+\Pi^{f}_{\rm vac})+\Pi_T^{\rm th}}, 
\hs{1.5cm}
 \Delta_L = \frac{-1}{p^2(1+\Pi^{f}_{\rm vac})+\Pi_L^{\rm th}}.
\ee
We note here that the product $e^{2}(\mu)\Delta_{T/L}$ is renormalization 
group invariant. As is well known, the theory has a Landau pole at the scale 
$\Lambda_{L}=\mu\,e^{6\pi^{2}/e^{2}(\mu)}$. The largest scale in the 
problem, either the temperature or the mass, has to be reasonably well 
below the Landau scale. This imposes a restriction on the allowed values 
of the coupling constant. Although the results are renormalization group 
invariant, in order to present them numerically we have to choose a scale. 
To facilitate a comparison between our results and the ones 
obtained for massless fermions in kinetic theory~\cite{Moore:2001fg}, we 
take $\mu=\mu_{\rm DR} = \pi e^{-\gamma_E}T$, the dimensional reduction
value for massless fermions.

The real part of the finite piece at zero temperature reads
\bea
 \re\,\Pi_{\rm vac}^f(P)=\frac{e^{2}}{12\pi^2}
 \Bigg\{&&\hm \left[\Theta(P^2-4m^2)+\Theta(-P^2)\right]
 \left(1+\frac{2m^2}{P^2}\right) 
 \beta(P)\ln\left|\frac{1-\beta(P)}{1+\beta(P)}\right|
 \nn \\ &&\hm 
 -2\Theta(4m^2-P^2)\Theta(P^2)\left(1+\frac{2m^2}{P^2}\right) 
 B(P) \arctan\frac{1}{B(P)}       \nn  \\ &&\hm 
 +\frac{4m^2}{P^2}-\ln\frac{m^2}{\mu^2}+\frac{5}{3}\Bigg\},
\eea
where
\be
 \beta(P) = \sqrt{1-\frac{4m^2}{P^2}},
 \;\;\;\;\;\;\;\;\;\;\;\;
 B(P) = \sqrt{\frac{4m^2}{P^2}-1}.
\ee

\begin{figure}[t]
 \centerline{\epsfig{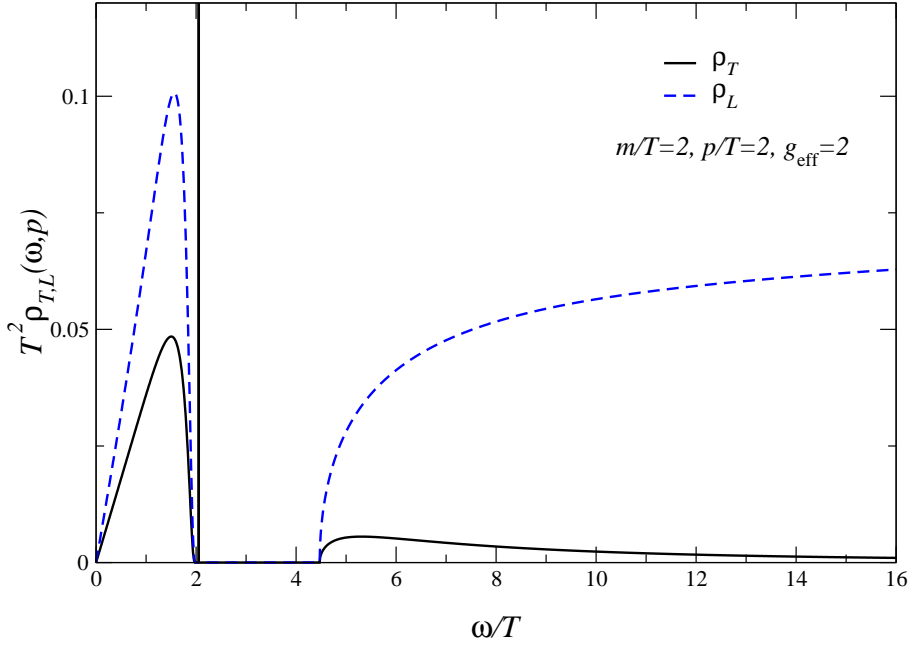}}
 \caption{Transverse and longitudinal spectral functions $\rho_{T,L}(\om,p)$ 
   for $p/T=2$, $m/T=2$ and $g_{\rm eff}=e=2$. For these parameters, the 
   transverse gauge boson is a stable quasiparticle at $\om_T/T=2.05$, 
   indicated with the vertical line. 
 }
 \label{fig:rhoTL}
\end{figure}

\begin{figure}[t]
 \centerline{\epsfig{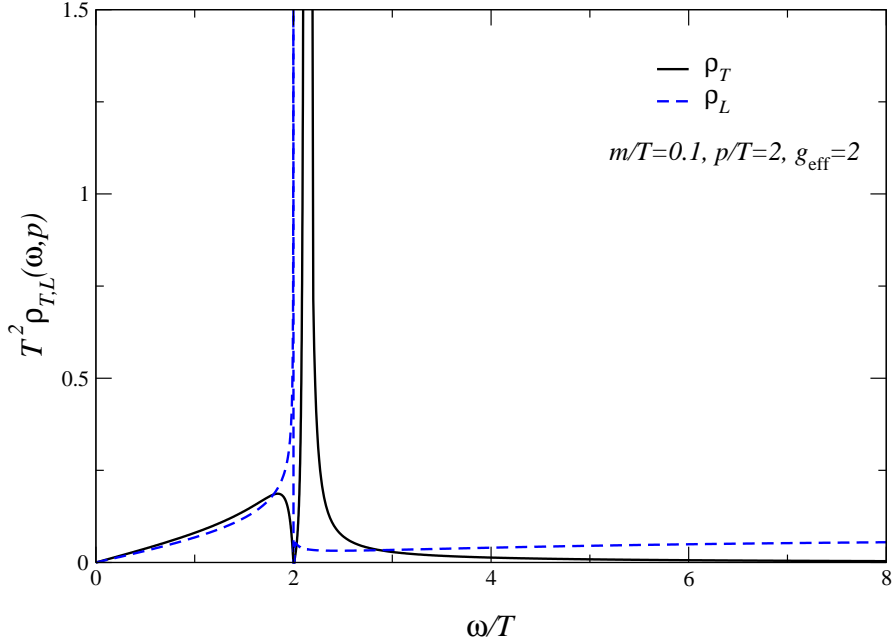}}
 \caption{As in Fig.\ \ref{fig:rhoTL}, with $m/T=0.1$. In this case 
 the transverse gauge boson is a resonance in the continuum at 
 $\om_T/T=2.13$. 
 }
 \label{fig:rhoTL-res}
\end{figure}

We now consider the thermal piece. For the real parts we find
\bea
 \re\,{\Pi^\mu_\mu}_{R,\rm th}(P) = &&\hm  
-\frac{4e^{2}}{\pi^2}\int_0^\infty 
 dk\,\frac{k^2}{\om_\kv}n_F(\om_\kv)  
 \left[1+\frac{P^2+2m^2}{8kp} 
 \ln\left|\frac{(k+p_+)(k+p_-)}{(k-p_+)(k-p_-)}\right|\right], \;\;\;\;
\\
 \re\,\Pi^{00}_{R,\rm th}(P) =  &&\hm -\frac{2e^{2}}{\pi^2}\int_0^\infty 
  dk\,\frac{k^2}{\om_\kv}n_F(\om_\kv)  
 \Bigg[1+\frac{P^2+4\om_\kv^2}{8kp}
 \ln\left|\frac{(k+p_+)(k+p_-)}{(k-p_+)(k-p_-)}\right|
 \nn  \\ &&\hm \hspace*{1.7cm}
 -\frac{p^0\om_\kv}{2pk}
 \ln\left|\frac{(P^2-2p^0\om_\kv-2pk)(P^2+2p^0\om_\kv+2pk)}
       {(P^2-2p^0\om_\kv+2pk)(P^2+2p^0\om_\kv-2pk)}\right|\Bigg],  
\eea
where $p_\pm=\half [p\pm p^0\beta(P)]$. The remaining 
one-dimensional integrals can be done numerically.
The imaginary parts, including the vacuum 
contribution, can be computed explicitly and read\footnote{
The arguments of the polylogarithmic functions $\li_n(z)$ are chosen
such that they lie between $-1$ and $0$ for positive $p^0$.}
\bea
&&\hm
 \im\,{\Pi^\mu_\mu}_{R,\rm th}(P)=\frac{e^{2}}{4\pi}(P^2+2m^2)\Bigg\{
 \Theta(P^2-4m^2)
 \left[\beta(P)+\frac{2T}{p}\ln\frac{1+e^{-\pbp/T}}{1+e^{-\pbm/T}}\right] 
 \nn \\  
 &&\hm\hspace*{5.8cm}
 +\Theta(-P^2) \frac{2T}{p} \ln \frac{1+e^{-\pbp/T}}{1+e^{\pbm/T}} \Bigg\},
\\
 &&\hm
 \im\,\Pi^{00}_{R,\rm th}(P) =
 -\frac{e^{2}T^2}{\pi}\Theta(P^2-4m^2)\Bigg\{
 \beta(P)\frac{P^2+2m^2}{12P^2}\frac{p^2}{T^2} 
 +\frac{m^2}{P^2}\frac{p}{T}\ln \frac{1+e^{-\pbp/T}}{1+e^{-\pbm/T}} 
 \nn \\ \nn 
 &&\hm+\beta(P)\left[ \li_2(-e^{-\pbp/T})+\li_2(-e^{-\pbm/T})\right]
 +\frac{2T}{p}\left[ \li_3(-e^{-\pbp/T}) - \li_3(-e^{-\pbm/T})\right]\Bigg\}
 \\ \nn 
 &&\hm-\frac{e^{2}T^2}{\pi}\Theta(-P^2)\Bigg\{
 \frac{m^2}{P^2}\frac{p}{T}\ln \frac{1+e^{-\pbp/T}}{1+e^{\pbm/T}}
 +\beta(P)\left[\li_2(-e^{-\pbp/T})-\li_2(-e^{\pbm/T})\right]  
 \\  &&\hm \hspace*{3cm}
 +\frac{2T}{p}\left[\li_3(-e^{-\pbp/T})-\li_3(-e^{\pbm/T})\right]\Bigg\},
\eea
where $\bar p_\pm = \half\left[p^0\pm p \beta(P)\right]$. 

The resulting transverse and longitudinal spectral densities are shown in 
Figs.\ \ref{fig:rhoTL} and \ref{fig:rhoTL-res}. The propagating transverse 
and longitudinal modes, $\om_{T/L}(p)$, are determined from the poles of 
the corresponding propagators in Eq.~(\ref{finiteprop}). For the specific 
parameters chosen in Figs.~\ref{fig:rhoTL} and \ref{fig:rhoTL-res}, the 
transverse gauge boson is stable in the first case and a resonance in the 
second. In both cases the longitudinal gauge boson does not propagate.


\subsection{Fermion}

The fermionic propagator is 
\be
 S(P)=\frac{-1}{i\tilde{\om}_{n}\gamma^0 -\gv\cdot\pv -m - \Sigma(P)}=
 -\int_{-\infty}^{\infty} \frac{d\om}{2\pi}\frac{\rho_F(\om,\pv)}{i\tilde{\om}_{n}-\om},
\ee
where $\tilde{\om}_{n}=(2n+1)\pi T$ ($n\in \mathbb{Z}$) is a fermionic 
Matsubara frequency and $\rho_F(\om,\pv)$ the spectral density of the fermion. 
The self energy can be decomposed as
\be
 \Sigma(P)=\gz\Sigma^{0}(P)+\gv\cdot\pvuni\,\Sigma^{s}(P)+\Sigma^{m}(P),
\ee 
such that the retarded electron propagator reads
\be
 S_R(P)=-\frac{ \gz\left[\po-\Sigma_R^0(P)\right]
 -\gv\cdot\pvuni\left[p+\Sigma_R^s(P)\right]+\left[m+\Sigma_R^m(P)\right]}
 {\left[\po-\Sigma_R^0(P)\right]^{2}-\left[p+\Sigma_R^s(P)\right]^{2}
 -\left[m+\Sigma_R^m(P)\right]^{2}}.
\ee
The poles of the retarded propagator at $\po=E_{\pv}-i\tw_{\pv}/2$
determine the properties of the quasiparticle excitations of the system. 
The self energy is
\be
 \label{se-sigma}
 \Sigma(P) = -\frac{e^2}{N_f} \si_R\gamma^\mu S(R)\gamma^\nu D_{\mu\nu}(R-P),
\ee
and we find
\be  \label{tw-def}
 E_{\pv} = \pm\om_\pv +\cO\left(\frac{1}{N_f}\right), 
 \hs{1cm} 
 \tw_\pv = -\left.\frac{\im\,\Sigma_R^{\rm sc}(P)}{\po} 
 \right|_{\po=\pm\om_\pv} + \cO\left(\frac{1}{N_f^2}\right),
\ee
with $\om_\pv = \sqrt{p^2+m^2}$ and 
\be \label{eqtrace}
 \Sigma_{R}^{\rm sc}(P) \equiv
 2\left[ \po\Sigma^0_R(P) + p\Sigma^p_R(P) + m\Sigma^m_R(P) \right]
 = \half\tr\,\Sigma_R(P)(\Pslash+m).
\ee
The retarded and advanced propagators then simplify to
\be
 S_{R/A}(P) = -\frac{\Pslash+m}{P^{2}-m^{2}-i\im\,\Sigma_{R/A}^{\rm sc}(P)},
\ee
and the corresponding spectral density is
\be
 \rho_F(P) = \left(\Pslash+m\right) \frac{ -2\im\,\Sigma_R^{\rm sc}(P)}
  {\left[p_0^2-\om_\pv^2\right]^2 + \left[\im\,\Sigma_R^{\rm 
sc}(P)\right]^2}.
\ee
In the large $N_{f}$ limit, this reduces to the spectral density of a free 
fermion,
\be
 \label{eqrhoF}
 \rho_F(P) = \left(\Pslash+m\right)\rho(P) 
 = \left(\Pslash+m\right)2\pi\,\sgn(\po)\delta(P^{2}-m^{2}). 
\ee
Nonetheless, whenever a pair of propagators with pinching poles is 
present, the large $N_{f}$ limit of the product of a retarded and an 
advanced propagator is
\be  \label{eqPP}
 S_R(P)\gamma^\mu S_A(P) = \left(\Pslash+m\right) 
 \gamma^\mu \left(\Pslash+m\right) \frac{\rho(P)}{2p^0\Gamma_\pv} + 
\cO(N_f^0). 
\ee

It remains to give the explicit expression for the thermal width. 
Combining Eqs.~(\ref{se-sigma}, \ref{eqtrace}) we find, after 
doing the Matsubara frequency sum using spectral representations, 
performing the analytic continuation and taking the trace, 
\bea
 &&\hm \im\,\tr\,\Sigma_R(P)(\Pslash+m) =
 -\frac{2e^2}{N_f} \int_R
 \left[ n_F(r^0) + n_B(r^0-p^0) \right] \rho(R) 
 \nn \\   &&\hm\times
 \left\{ 2\rho_T(R-P ) \left[r^0p^0-m^2-\rv\cdot\kvuni\,\pv\cdot\kvuni\right]
 +\rho_L(R-P) \left[ r^0p^0+m^2+\rv\cdot\pv \right] \right\},
\eea
where $\kv=\rv-\pv$. Since only the spectral density of the gauge boson 
propagator contributes to the thermal width, this is explicitly 
independent of the gauge fixing parameter.

We proceed by introducing $k = |\rv -\pv|$ as
\be  \label{eqkrp}
 1 = \int_0^\infty dk\,\delta(k-|\rv -\pv|) = \int_{|r-p|}^{r+p}
 dk\,\frac{k}{rp}\delta(\cos \theta_{pr} -z_{pr}),
\ee
where $\cos \theta_{pr} = \pvuni\cdot\rvuni$ is the cosine of the angle
between $\pv$ and $\rv$ and
\be \label{eqzpr}
 z_{pr} = \frac{r^2+p^2-k^2}{2rp}. 
\ee 
The integral over $r^0$ is performed using $\rho(R)$ and the one over 
$\theta_{pr}$ using the delta function introduced above. After that, the 
final result for the thermal width reads
\bea 
 \Gamma_\pv = &&\hm \frac{e^2}{8N_f\pi^2p\,\om_\pv} \int_0^\infty 
 dr\,\frac{r}{\om_\rv} \int_{|r-p|}^{r+p} dk\,k      
 \nn  \\   
 &&\hm \times \Big\{ \left[ n_F(\om_\rv)+n_B(\om_\rv+\om_\pv) \right] 
 \left[ 2c_T^+\rho_T(\om_\rv+\om_\pv,k) - c_L^+\rho_L(\om_\rv+\om_\pv,k)\right] 
 \nn \\   &&\hm
 \label{eqthermalwidth}
 -\left[ n_F(\om_\rv)+n_B(\om_\rv-\om_\pv) \right] 
 \left[ 2c_T^-\rho_T(\om_\rv-\om_\pv,k)
  -c_L^-\rho_L(\om_\rv-\om_\pv,k) \right] \Big\}, 
\eea 
with
\be
 \label{eqcTcL}
 c_T^\pm = \pm\om_\rv\om_\pv + \frac{pr}{k^2}(r-pz_{pr})(p-rz_{pr}) + m^2, 
 \hs{1cm}
 c_L^\pm = \mp\om_\rv\om_\pv+prz_{pr}+m^2.
\ee
It should be noted that the thermal width as it is written here is not 
defined, due to the divergent contribution from soft quasistatic 
transverse gauge bosons \cite{smilga,Blaizot:1996az}. However, in 
the application 
to transport coefficients this contribution cancels against part 
of the ladder diagrams. This has been analyzed in detail in the 
weak coupling limit in Refs.\ \cite{ValleBasagoiti:2002ir,Aarts:2002tn}.


\section{Shear viscosity}
\label{shear}

\begin{figure}[t]
 \centerline{\epsfig{figure=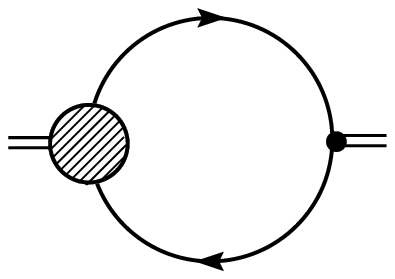,height=1.9cm}}
 \caption{Effective one-loop diagram contributing to the shear viscosity.}
 \label{fig:eta}
\end{figure}

We are now ready to compute the shear viscosity. It is obtained from the 
effective one-loop diagram (see Fig.\ \ref{fig:eta})  
\be
 G_{\pi\pi}(Q)=-\si_P\tr\,S(P+Q)\Gamma^{ij}(P+Q,P)S(P)\Gamma^{ij}_0(\pv),
\ee
where $Q=(i\om_q,\vecnul)$ is the momentum that enters from the left. 
The coupling between the external operator and two fermionic fields is 
$\Gamma_0^{ij}(\pv)=p\,T^{ij}[\gv,\pvuni]$, with
\be
 T^{ij}[\gv,\pvuni]= 
\half\left( \gamma^{i}\puni^{j}+\gamma^{j}\puni^{i} \right)
 -\frac{1}{3}\delta^{ij}\gv\cdot\pvuni.
\ee
In order to carry out the sum over Matsubara frequencies, we use the 
method described in Ref.~\cite{ValleBasagoiti:2002ir}. From the analytic 
properties of the propagators and the structure of the integral equation, 
it follows that the effective vertex $\Gamma^{ij}(P+Q,P)$ has cuts along 
$\im(\po)=0$ and $\im(\po+\qo)=0$. As is usually the case 
\cite{ValleBasagoiti:2002ir,Aarts:2002tn,Aarts:2004sd}, in the pinching 
pole limit only one particular analytic continuation of the effective 
vertex is required. At leading order in the $1/N_f$ expansion we arrive at
\be
 \lim_{q^0\to 0}\rho_{\pi\pi}(q^0,\zv)=-2q^0 N_f\int_P n_F'(p^0)\re\,\tr\,
 S_R(P) \Gamma^{ij}(p^0+i0,p^0-i0;\pv) S_A(P)\Gamma^{ij}_0(\pv).
\ee
Using now Eq.\ (\ref{eqPP}) for the product of the retarded and the 
advanced Green functions, the viscosity reads
\be
 \eta=-\frac{N_f}{20}\int_P n_F'(p^0)\frac{\rho(P)}{p^0\Gamma_\pv}  
 \re\, \tr\, (\Pslash+m) \Gamma^{ij}(p^0+i0,p^0-i0;\pv)(\Pslash+m) 
 \Gamma^{ij}_0(\pv).
\ee
Due to the pinching poles, the momentum in the loop is forced onshell: 
$p^0=\pm\om_\pv$. We may therefore decompose $\Pslash+m$ into spinors 
according to
\be
\Pslash+m \Big|_{p^0 = \om_\pv} = \sum_{\lambda=1}^2 u_\lambda(\pv)\bar 
u_\lambda(\pv), 
\;\;\;\;\;\;\;\;\;\;\;
\Pslash+m \Big|_{p^0 = -\om_\pv} = -\sum_{\lambda=1}^2 v_\lambda(-\pv)\bar 
v_\lambda(-\pv),
\ee
and associate the spinors with the effective vertex $\tw^{ij}$ as
\bea
\nn &&\hm 
\re\,
\bar{u}_{\lambda}(\pv)\tw^{ij}(\om_{\pv},\om_{\pv};\pv)u_{\lambda}(\pv) 
\\
&&\hm =
\re\, 
\bar{v}_{\lambda}(-\pv)\tw^{ij}(-\om_{\pv},-\om_{\pv};\pv)v_{\lambda}(-\pv)
= 2p^{2}T^{ij}[\pvuni,\pvuni]\cD(p).
\label{eqspinor}
\eea
An equivalent expression for the resulting scalar vertex $\cD(p)$ 
is
\be
 \label{eqdef}
 \cD(p)=\frac{3}{8p^{2}}\re\,\tr\, 
 \Gamma^{ij}(p^0+i0,p^0-i0;\pv)(\Pslash+m)T^{ij}[\pvuni,\pvuni]
 \Big|_{p^0=\pm\om_\pv}.
\ee
The normalization is such that for the bare vertex $\tw^{ij}_0(\pv)$ 
this yields $\cD(p)=1$. It is then straightforward to obtain\footnote{
A different route to arrive at these expressions is as follows. Inspection 
of the integral equation shows that in the
pinching pole limit the effective vertex remains linear in the
$\gamma$-matrices and the identity and can be taken traceless. It can 
therefore be decomposed as
\[
\Gamma^{ij}(P) = A(P)pT^{ij}[\gv,\pvuni]
+ T^{ij}[\pvuni,\pvuni]
\left\{ B(P) \gv\cdot\pv + C(P) \gamma^0p^0 + D(P) m \right\},
\]
with 4 independent scalar functions $A,B,C$ and $D$.
However, an analysis of the integral equation with this decomposition 
shows that only one linear combination of those four functions 
enters in the final scalar equation, namely 
\[
\cD(p) = \frac{1}{p^2}\re \left\{ p^2\left[A(p) + B(p) + C(p)\right]
+ m^2 \left[C(p) + D(p)\right] \right\} \Big|_{p^0=\pm\om_\pv},
\]
as expected from the spinor decomposition.
} 
\be  \label{eqetachi}
 \eta=-\frac{4N_f}{15}\int_\pv\frac{p^2}{\om_\pv} n_F'(\om_\pv)\chi(p) 
 = -\frac{2N_f}{15\pi^2}
 \int_0^\infty dp\,\frac{p^4}{\om_\pv}n_F'(\om_\pv)\chi(p),
\ee
where 
\be
 \label{eqchishear}
 \chi(p)=\frac{p^2}{\om_\pv} \frac{\cD(p)}{\Gamma_\pv}.
\ee


\subsection{Integral equation}

From the integral equation (\ref{ie}), we now obtain an integral equation 
for the effective vertex $\cD(p)$. To do the frequency sums we proceed as 
before (see Ref.~\cite{Aarts:2004sd} for more details in a similar 
computation). We find
\bea
 \cD(p)=1&&\hm+\int_R\left[n_B(r^0-p^0)+n_F(r^0)\right] 
 \frac{r^2}{p^2}P_2(\pvuni\cdot\rvuni)\frac{\rho(R)}{2r^0\Gamma_\rv}\cD(r) 
 \Lambda_{\rm line}(R,P)
 \nn \\
 &&\hm+\int_{R}\left[n_B(r^0-p^0)+n_F(r^0)\right] 
 \frac{r^2}{p^2} P_2(\pvuni\cdot\rvuni)\frac{\rho(R)}{2r^0\Gamma_\rv}\cD(r) 
 \Lambda_{\rm box\, 1}(R,P)
 \nn \\ 
 \label{eqDeta}
 &&\hm+\int_{R}\left[n_B(r^0+p^0)+n_F(r^0)\right] 
 \frac{r^2}{p^2} P_2(\pvuni\cdot\rvuni)\frac{\rho(R)}{2r^0\Gamma_\rv}\cD(r) 
 \Lambda_{\rm box\, 2}(R,P),
\eea
where here and below $p^0=\pm\om_\pv$. 
The three contributions arise from the line diagram and the two box 
diagrams respectively. The second Legendre polynomial $P_2(x) = 
(3x^2-1)/2$ originates from
\be
T^{ij}[\rvuni,\rvuni]T^{ij}[\pvuni,\pvuni]
 = \frac{2}{3}P_2(\pvuni\cdot\rvuni).
\ee
The three kernels read
\bea  \label{furry}
 \Lambda_{\rm line}(R,P) = &&\hm 
 \frac{e^2}{2N_f}\tr\left[\gamma^\mu(\Rslash+m)\gamma^\nu(\Pslash+m)\right]
 \rho_{\mu\nu}(R-P), \nn \\
 \Lambda_{\rm box\, 1}(R,P) = &&\hm 
 \frac{e^4}{2N_f}\int_{L}  \left[n_F(l^0-r^0)-n_F(l^0-p^0)\right]
 D^R_{\rho\mu}(L)D^A_{\nu\sigma}(L)
 \nn \\ &&\hm \times
 \tr\left[\gamma^\mu(\Rslash+m)\gamma^\nu\rho_F(R-L)\right]
 \tr\left[ \gamma^\rho \rho_F(P-L)\gamma^\sigma (\Pslash+m) \right],  \nn \\
 \Lambda_{\rm box\, 2}(R,P) = &&\hm
 \frac{e^4}{2N_f}\int_{L}
 \left[n_F(l^0+r^0)-n_F(l^0-p^0)\right]
 D^R_{\rho\mu}(L)D^A_{\nu\sigma}(L)  
 \nn \\ &&\hm  \times
 \tr\left[\gamma^\nu(\Rslash+m)\gamma^\mu\rho_F(R+L)\right]
 \tr\left[ \gamma^\rho \rho_F(P-L)\gamma^\sigma (\Pslash+m) \right].
\eea 
Using Eq.\ (\ref{eqrhoF}) for the fermionic spectral functions, it is easy 
to see that under a change of variables $R\to -R$, the contribution from 
the second box diagram becomes identical to the first one. We can then 
write 
\be
 \cD(p) = 1 +\int_R \left[n_B(r^0-p^0)+n_F(r^0)\right] \frac{r^2}{p^2} 
 P_2(\pvuni\cdot\rvuni)\frac{\rho(R)}{2r^0\Gamma_\rv} \cD(r) \Lambda(R,P),
\ee
with
\bea  \label{rung}
 \Lambda(R,P)=&&\hm\frac{e^2}{2N_f}
 \tr\left[\gamma^\mu(\Rslash+m)\gamma^\nu(\Pslash+m)\right]
 \rho_{\mu\nu}(R-P) \nn  \\ \nn
 &&\hm+  \frac{e^4}{N_f}\int_L \left[n_F(p^0-l^0)-n_F(r^0-l^0)\right]
 \rho(P-L)\rho(R-L) D^R_{\rho\mu}(L) D^A_{\nu\sigma}(L) 
 \\ &&\hm \times 
 \tr\left[\gamma^\mu (\Rslash+m) \gamma^\nu (\Rslash-\Lslash+m)\right]  
 \tr\left[\gamma^\rho (\Pslash-\Lslash+m) \gamma^\sigma (\Pslash+m)\right].
\eea
In terms of the function $\chi(p)$ defined in Eq.\ (\ref{eqchishear}) the 
integral equation reads
\be  \label{ie-chi}
 \om_\pv\Gamma_\pv\chi(p) = p^2 + 
 \half\int_R\left[n_B(r^0-p^0)+n_F(r^0)\right] 
 P_2(\pvuni\cdot\rvuni)\frac{\om_\rv}{r^0} \chi(r) \rho(R) \Lambda(R,P),
\ee
where $p^0=\pm\om_\pv$. Upon solving it for $\chi(p)$, we obtain the 
shear viscosity from Eq.~(\ref{eqetachi}).


\subsection{Variational approach}

Since the integral equation looks prohibitively difficult to solve 
analytically, we proceed to formulate it as a variational problem, which 
gives a convenient formulation for finding a numerical 
solution~\cite{Arnold:2000dr,Aarts:2004sd}.

After multiplying Eq.\ (\ref{ie-chi}) with
\be \label{eqmult1}
 \frac{p^2}{\om_\pv}n_F'(\om_\pv),
\ee
the integral equation can be written as
\be   \label{eqint1}
 {\cal F}(p)\chi(p) = {\cal S}(p)  + \int_0^\infty dr\, {\cal H}(p,r)\chi(r),
\ee
with
\be
 {\cal F}(p) =  p^2 n_F'(\om_\pv) \Gamma_\pv,
 \;\;\;\;\;\;\;\;\;\;\;\;
 {\cal S}(p) = \frac{p^4}{\om_\pv}n_F'(\om_\pv),
\ee
and a symmetric kernel, ${\cal H}(p,r)={\cal H}(r,p)$, whose explicit form 
is presented below. Since ${\cal H}$ is symmetric,
Eq.\ (\ref{eqint1}) can be derived by extremizing the functional
\be \label{Q}
 Q[\chi]=\int_0^{\infty}dp\left[{\cal S}(p)\chi(p)-\half{\cal F}(p)\chi^2(p)
 +\half\int_0^\infty dr\, {\cal H}(p,r)\chi(r)\chi(p) \right].
\ee
The viscosity is then given by the extremum of the functional
\be
 \eta=-\frac{4N_f}{15\pi^2}Q[\chi=\chi_{\rm ext}].
\ee
 In the rest of this section we explicitly evaluate ${\cal H}(p,r)$. 

We separately compute the single line diagram and the box diagram, ${\cal 
H} = {\cal H}_{\rm line} + {\cal H}_{\rm box}$. We start with the diagram 
containing the single line. Proceeding as we did in the calculation of 
the thermal width, we arrive at\footnote{In Ref.~\cite{Aarts:2004sd}, the 
first term in Eqs.~(5.17) and (5.32) is written with the wrong sign.}
\bea
 \label{eqHline}
 {\cal H}_{\rm line}(p,r)=&&\hm-\frac{e^2}{8N_f\pi^2} n_F'(\om_{\pv})
 \frac{p}{\om_\pv} \frac{r}{\om_\rv}\int_{|r-p|}^{r+p} dk\,k P_2(z_{pr})
 \\  &&\hm \times 
 \Big\{ \left[ n_F(\om_\rv)+n_B(\om_\rv+\om_\pv) \right] 
 \left[2c_T^+\rho_T(\om_\rv+\om_\pv,k)-c_L^+\rho_L(\om_\rv+\om_\pv,k)\right] 
 \nn \\ 
 \nn &&\hm 
 + \left[ n_F(\om_\rv)+n_B(\om_\rv-\om_\pv) \right]
 \left[2c_T^-\rho_T(\om_\rv-\om_\pv,k)-c_L^-\rho_L(\om_\rv-\om_\pv,k)\right]
 \Big\}.
\eea
The coefficients $c^\pm_{T/L}$ were already defined in Eq.\ 
(\ref{eqcTcL}). The contribution from the line diagram is $\xi$ 
independent for the same reason as the thermal width. Using similar 
properties of the distribution functions as discussed in Ref.\ 
\cite{Aarts:2004sd}, it is straightforward to verify that 
${\cal H}_{\rm line}$ is symmetric under exchange of $p$ and $r$.

For the contribution from the box diagrams we work out the traces and 
contractions in Eq.\ (\ref{rung}). After a bit of algebra we 
find\footnote{In Appendix B of Ref.~\cite{Moore:2001fg}, the factor 
$\re\,\Delta^{R}_{T}\Delta^{A}_{L}$ appears written erroneously as 
$|\Delta^{R}_{T}\Delta^{A}_{L}|$.} 
\bea  
 \label{trace}
 &&\hm
 D^R_{\rho\mu}(L) D^A_{\nu\sigma}(L) 
 \tr\left[\gamma^\mu(\Rslash+m)\gamma^\nu(\Rslash-\Lslash+m)\right] 
 \tr\left[\gamma^\rho(\Pslash-\Lslash+m)\gamma^\sigma(\Pslash+m)\right]
 \nn \\
 = &&\hm 
 \Delta^{R}_{L}(L)\Delta^{A}_{L}(L)
 \left(2r_0^2-r^0l^0-\rv\cdot\lv\right)
 \left(2p_0^2-p^0l^0-\pv\cdot\lv\right)
 \nn \\
 + &&\hm 
 2\re[\Delta^{R}_{T}(L)\Delta^{A}_{L}(L)]
 \left(2r^0-l^0 \right)\left(2p^0-l^0 \right)
 \left(\pv\cdot\rv-\pv\cdot\lvuni\,\rv\cdot\lvuni\right) 
 \nn \\
 + &&\hm 2 \Delta^{R}_{T}(L)\Delta^{A}_{T}(L) 
 \Big[R\cdot L\,P\cdot L
 -P\cdot L\left(r^2-(\rv\cdot\lvuni)^2\right)
 -R\cdot L\left(p^2-(\pv\cdot\lvuni)^2\right)
  \nn \\ &&\hm \hspace*{2.9cm} 
 +2\left(\pv\cdot\rv-\pv\cdot\lvuni\,\rv\cdot\lvuni\right)^2\Big]
 \nn \\
 + &&\hm  \frac{1}{l^4} \left( 2R\cdot L-L^2 \right)
 \left( 2P\cdot L-L^2\right)   
 \Big[2\xi
 \left(\pv\cdot\rv-\pv\cdot\lvuni\,\rv\cdot\lvuni\right)
 \re\Delta^{R}_{T}(L)
 \nn \\ &&\hm \hspace*{2.9cm}
 +2\xi r^0p^0\,\re\Delta^{R}_{L}(L)
 +\frac{\xi^{2}}{l^4}R\cdot L\,P\cdot L\Big].
\eea
The terms that depend on the gauge fixing parameter $\xi$ are 
proportional to $2P\cdot L-L^{2}$. These terms are accompanied by 
the Dirac delta function $\rho(P-L)$ (see Eq.~(\ref{rung})), which for 
onshell momentum $P$ causes this factor to vanish. Since this was the last 
piece that depended on the gauge boson propagator, we find that the viscosity 
is gauge fixing independent, as it should obviously be. 

To proceed further, we consider the 8-dimensional integral over $R$ and $L$ 
in Eqs.\ (\ref{ie-chi},\ref{rung}). The cosine of the angle between 
$\pv$ and $\lv$ is denoted as $\cos \theta_{pl}$, between 
$\rv$ and $\lv$ as $\cos\theta_{rl}$, and the azimuthal angle between the 
$\pv,\lv$ plane and the $\rv,\lv$ plane as $\phi$. We specialize to $p^0=
\om_\pv$. The 8-dimensional 
integral can then be written as
\be
 \frac{2\pi}{(2\pi)^8}
 \int_0^\infty dr\,r^2 \int_{-\infty}^\infty dr^0
 \int_0^\infty dl\,l^2 \int_{-\infty}^\infty dl^0
 \int_{-1}^1 d\!\cos \theta_{pl} \int_{-1}^1 d\!\cos\theta_{rl}
 \int_0^{2\pi}d\phi.
\ee
 The integration over $\cos \theta_{pl}$ will be performed using the delta 
functions in $\rho(P-L)$, the one over $\cos \theta_{rl}$ using 
$\rho(R-L)$ and that over $r^0$ using $\rho(R)$. The product of the three 
spectral functions yields a set of constraints, since
\bea
 \rho(R)\rho(P-L)\rho(R-L)\Big|_{p^0=\om_\pv} \!\!\!\! 
 \sim &&\hm \sum_{s_i=\pm}
 \delta(r^0+s_1\om_\rv)\delta(\om_\pv-l^0+s_2\om_{\pv-\lv})
 \delta(r^0-l^0-s_3\om_{\rv-\lv}) 
 \nn \\
 \sim &&\hm\sum_{s_i=\pm}
 \delta(\om_\pv+s_1\om_\rv+s_2\om_{\pv-\lv}+s_3\om_{\rv-\lv}).
\eea
Out of the eight combinations, only three can contribute for kinematical
reasons, namely those corresponding to $2\leftrightarrow 2$ processes. We
treat these three cases separately and write
\be
 {\cal H}_{\rm box} = {\cal H}_{\rm box}^{(1)} +
 {\cal H}_{\rm box}^{(2)} + {\cal H}_{\rm box}^{(3)}.
\ee

\begin{enumerate}

\item $(s_1, s_2, s_3) = (-,+, -)$.
The cosines are $\cos \theta_{pl} = z_{pl}$, $\cos \theta_{rl} =
z_{rl}^-$, where
\be
 z_{pl} = \frac{l^2-l_0^2}{2pl} +\frac{\om_\pv l^0}{pl},
 \;\;\;\;\;\;\;\;
 z_{rl}^{s_1} = \frac{l^2-l_0^2}{2rl} -s_1\frac{\om_\rv l^0}{rl}.
\ee
The constraints from the spectral functions can be satisfied provided
\be
 l^0> \sqrt{l^2+4M^2}, \;\;\;\;\;\;\;\; |l_-| < p,r < |l_+|,
\ee
where we use again the notation
\be
 l_\pm = \half \left[l\pm l^0\beta(L)\right], \;\;\;\;\;\;\;\;
 \beta(L) = \sqrt{1-\frac{4M^2}{L^2}}.
\ee
The angle $\phi$ appears both in expression (\ref{trace}) and in 
$P_2(\pvuni\cdot\rvuni)$ in Eq.\ (\ref{ie-chi}), since 
\be
 \pvuni\cdot\rvuni=\sin\theta_{pl}\sin\theta_{rl}\cos\phi
 +\cos\theta_{pl}\cos\theta_{rl}.
\ee
Performing the integral over $\phi$ yields an expression of the form
\be
 c_{LL} \left|\Delta_L^R(L)\right|^2+c_{TL}\re\,\Delta_T^R(L)\Delta_L^A(L)+
 c_{TT} \left|\Delta_T^R(L)\right|^2,
\ee
with
\bea
 c_{LL} = &&\hm \frac{1}{4} \left[ L^2 + 4\om_{\pv}(\om_{\pv}-\lo) \right]
 \left[ L^2 + 4\om_{\rv}(\om_{\rv}+s_{1}\lo) \right]
 \phi_0(\cos\theta_{pl},\cos\theta_{rl}),
 \nn \\
 c_{TL} = &&\hm p\,r(2\om_{\pv}-\lo)\left(-2s_1\om_\rv-l^0\right)
 \phi_1(\cos\theta_{pl},\cos\theta_{rl}),
 \nn \\
 c_{TT} = &&\hm
 \frac{L^2}{2}\left( L^2 - 2p^2\sin^2\theta_{pl}
 -2r^2\sin^2\theta_{rl} \right)
 \phi_0(\cos\theta_{pl},\cos\theta_{rl})
 \nn \\ &&\hm 
 + 4p^{2}r^{2}\phi_{2}(\cos\theta_{pl},\cos\theta_{rl}),
\eea
where
\be
 \phi_{n}(\cos\theta_{pl},\cos\theta_{rl})=\int_{0}^{2\pi}\frac{d\phi}{2\pi}
 P_{2}(\cos\theta_{pr})(\cos\theta_{pr}-\cos\theta_{pl}\cos\theta_{rl})^{n}.
\ee
The explicit expressions we need are
\bea
 \phi_0(x,y) = &&\hm P_2(x)P_2(y),
 \nn\\
 \phi_1(x,y) = &&\hm \frac{3}{2}xy(x^2-1)(y^2-1),
 \nn\\
 \phi_2(x,y) = &&\hm \frac{1}{16}(x^2-1)(y^2-1)(5-9x^2-9y^2+21x^2y^2). 
\eea

Multiplying the resulting expression with Eq.\ (\ref{eqmult1}) we can read
off the first contribution to ${\cal H}(p,r)$ from the box diagram:
\bea
 &&\hm{\cal H}_{\rm box}^{(1)}(p,r)=
 \frac{1}{8\pi^3}\frac{e^4}{N_f} \frac{p}{\om_\pv}\frac{r}{\om_\rv}
 n_F'(\om_\pv)\left[n_B(\om_\rv-\om_\pv) + n_F(\om_\rv) \right]
 \nn \\ 
 &&\hm\times\int_0^\infty dl\int_{\sqrt{l^2+4M^2}}^\infty dl^0 
 \left(c_{TT} \left|\Delta_T^R(L)\right|^2 + 
 c_{LL}\left|\Delta_L^R(L)\right|^2
 +c_{TL} \re\,\Delta_T^R(L)\Delta_L^A(L)\right) 
 \nn \\ 
 &&\hm \times\left[n_F(\om_\pv-l^0)-n_F(\om_\rv-l^0)\right]
 \Theta(p-|l_-|)\Theta(|l_+|-p)\Theta(r-|l_-|)\Theta(|l_+|-r).
 \nn \\ &&
\eea

\item $(s_1,s_2,s_3) = (-,-,+)$.
The cosines are $\cos \theta_{pl} = z_{pl}$, $\cos \theta_{rl} =
z_{rl}^-$ and the constraints are
\be
l_0^2< l^2, \;\;\;\;\;\;\;\; p > |l_+|, \;\;\;\;\;\;\;\; r > |l_+|.
\ee
Therefore the second contribution reads
\bea
&&\hm \nn
{\cal H}_{\rm box}^{(2)}(p,r) =
\frac{1}{8\pi^3} \frac{e^4}{N_f}\frac{p}{\om_\pv}\frac{r}{\om_\rv}
n_F'(\om_\pv)\left[n_B(\om_\rv-\om_\pv) + n_F(\om_\rv) \right]
 \\ && \times\nn
\int_0^\infty dl \int_{-l}^l dl^0
\left(
c_{TT} \left|\Delta_T^R(L)\right|^2 + 
c_{LL} \left|\Delta_L^R(L)\right|^2 + 
c_{TL} \re\, \Delta_T^R(L)\Delta_L^A(L)
\right) 
 \\ && \times
 \left[n_F(\om_\pv-l^0)-n_F(\om_\rv-l^0)\right]
\Theta\left(p-|l_+|\right)\Theta\left(r-|l_+|\right).
\eea

\item $(s_1,s_2,s_3) = (+,-,-)$.
In this case the cosines are $\cos \theta_{pl} = z_{pl}$, $\cos
\theta_{rl} = z_{rl}^+$, with the constraints
\be
l_0^2< l^2, \;\;\;\;\;\;\;\; p > |l_+|, \;\;\;\;\;\;\;\; r > |l_-|.
\ee
The third contribution then reads
\bea
&&\hm \nn
{\cal H}_{\rm box}^{(3)}(p,r) =
\frac{1}{8\pi^3} \frac{e^4}{N_f}\frac{p}{\om_\pv}\frac{r}{\om_\rv}
n_F'(\om_\pv)\left[n_B(\om_\rv+\om_\pv) + n_F(\om_\rv) \right]
 \\ && \times\nn
\int_0^\infty dl \int_{-l}^l dl^0 
\left(
c_{TT} \left|\Delta_T^R(L)\right|^2 + 
c_{LL} \left|\Delta_L^R(L)\right|^2 + 
c_{TL} \re\, \Delta_T^R(L)\Delta_L^A(L) 
\right) 
 \\ && \times
\left[n_F(l^0+\om_\rv)-n_F(l^0-\om_\pv)\right]
\Theta\left(p-|l_+|\right)\Theta\left(r-|l_-|\right).
\eea

\end{enumerate}
Again, one can verify that ${\cal H}_{\rm box}$ is symmetric under 
exchange of $p$ and $r$, which allows for the integral equation 
to be derived from the functional $Q$.


\section{Electrical conductivity}
\label{conductivity}

The calculation of the electrical conductivity goes along the same 
lines as above. 
The coupling between two fermion fields and the external 
operator is simply $\gamma^i$. The correlator we need reads then
\be
 G_{jj}(Q) = - q_f^2\si_P\tr\,S(P+Q)\Gamma^{i}(P+Q,P)S(P)\gamma^{i}.
\ee
Proceeding as we did for the shear viscosity, we introduce a scalar 
vertex as 
\be
 \re\, \bar{u}_{\lambda}(\pv)\tw^{i}(\om_{\pv},\om_{\pv};\pv)u_{\lambda}(\pv)
 = \re\,  \bar{v}_{\lambda}(-\pv) \tw^{i}(-\om_{\pv},-\om_{\pv};\pv) 
 v_{\lambda}(-\pv)
 = 2p^{i}\cD(p),
\ee
or through the equivalent expression 
\be
 \cD(p) = \frac{1}{4p}\re\,\tr\, 
 \Gamma^{i}(p^0+i0,p^0-i0;\pv)(\Pslash+m)\puni^{i}\Big|_{p^0=\pm\om_\pv}.
\ee
We then find for the conductivity
\be
  \sigma=-\frac{4q_f^2N_f}{3}\int_\pv\frac{p}{\om_\pv}n_F'(\om_\pv)\chi(p)
  = -\frac{2q_f^2N_f}{3\pi^2}\int_0^\infty dp\,
  \frac{p^3}{\om_\pv}n_F'(\om_\pv)\chi(p).
\ee
with
\be
 \label{eqchiconduct}
 \chi(p) = \frac{p}{\om_\pv} \frac{\cD(p)}{\Gamma_\pv}.
\ee

The integral equation is actually simpler, since the contributions from 
the diagrams with the box rungs cancel each other. This is a consequence 
of Furry's theorem. It can be seen explicitly as follows. The 
integral equation for the conductivity has a similar form as Eq.\ 
(\ref{eqDeta}), but with $P_2(\pvuni\cdot\rvuni)$ replaced by 
$P_1(\pvuni\cdot\rvuni)=\pvuni\cdot\rvuni$. Under the change of variables 
$R\to -R$ mentioned below Eq.\ (\ref{furry}), this factor is now odd and 
the second box diagram cancels exactly the first one.
The integral equation reads therefore
\be
 \Gamma^{i}(P+Q,P) = \gamma^i + \frac{e^2}{N_f}\si_R 
 \gamma^\nu S(R+Q) \Gamma^{i}(R+Q,R) S(R) \gamma^\mu D_{\mu\nu}(R-P).
\ee
Proceeding as before, we arrive at
\be  \label{eqchicon}
 \om_\pv\Gamma_\pv\chi(p) = p + \frac{1}{2}\int_R 
 \left[n_F(r^0)+n_B(r^0-\om_\pv)\right] \pvuni\cdot\rvuni
 \frac{\om_\rv}{r^0} \chi(r) \rho(R)\Lambda(R,P),
\ee
with the kernel
\be
 \Lambda(R,P) = \frac{e^2}{2N_f}
 \tr\left[(\Rslash+m)\gamma^\mu(\Pslash+m)\gamma^\nu\right] 
 \rho_{\mu\nu}(R-P).
\ee
Upon multiplication with
\be  \label{eqmult2}
 \frac{p^2}{\om_\pv} n_F'(\om_\pv),
\ee
the integral equation can be cast in the form of Eq.~(\ref{eqint1}), where 
${\cal F}(p)$ is the same as above, but 
\be
 {\cal S}(p) = \frac{p^3}{\om_\pv}n_F'(\om_\pv).
\ee
 The kernel ${\cal H}$ receives only one contribution, ${\cal H}_{\rm 
line}$, which can be obtained from Eq.\ (\ref{eqHline}) by the replacement 
$P_2(z_{pr}) \to P_1(z_{pr})=z_{pr}$. The electrical conductivity is given 
by
\be
 \sigma = -\frac{4q_f^2N_f}{3\pi^2}Q[\chi=\chi_{\rm ext}].
\ee


\subsection{Ward identity}

The Ward identity relates the fermion-gauge boson vertex and the fermion 
self energy. We can explicitly verify that the integral equation which 
determines the fermion-gauge boson vertex is consistent with the Ward 
identity. We follow closely our analysis in the weakly coupled limit 
\cite{Aarts:2002tn}.
 
The abelian Ward identity for the fermion-gauge boson vertex reads
\be
 Q_{\mu}\Gamma^{\mu}(P+Q,P) = S^{-1}(P)-S^{-1}(P+Q).
\ee
After performing the required analytic continuation and taking $\qv=0$ 
this becomes
\be
 \qo\tw^{0}(\po+\qo+i0^{+},\po-i0^{+};\pv) = 
 \qo\gz+\Sigma^{A}(P)-\Sigma^{R}(P+Q).
\ee
In terms of 
\be
 \label{eqmathD}
 \mathfrak{D}(p)= \lim_{q^0\to 0} q^0 \frac{1}{4\po}
\tr \left[ \tw^{0}(\po+\qo+i0^{+},\po-i0^{+};\pv)(\Pslash +M)\right]
 \Big|_{\po=\pm\om_{p}},
\ee
the Ward identity in the pinching pole limit reads then
\be
 \label{eqWI}
 \mathfrak{D}(p)=i\tw_{\pv}.
\ee
The integral equation for the zero'th component of the effective 
vertex is
\be \label{eqint0}
 \tw^{0}(P+Q,P) = \gz+\frac{e^{2}}{N_{f}}\si_{R}
 \gamma^{\mu}S(R+Q)\tw^{0}(R+Q,R)S(R)\gamma^{\nu}D_{\mu\nu}(R-P).
\ee
After performing the sums, the analytic continuation and taking 
appropriate limits as in Eq.\ (\ref{eqmathD}) above,
we find
\bea
 &&\hm\hm\hm\mathfrak{D}(p)=\frac{2e^{2}}{N_f}\int_{R} 
 \left[n_{B}(\ro-\po)-n_{F}(\ro)\right]\frac{\rho(R)}{2\ro\tw_{\rv}} 
 \mathfrak{D}(r)\frac{\ro}{\po}
 \nn \\
 &&\hm\hm\times
 \left\{2\rho_{T}(R-P)\left[\ro\po-m^2-\kv\cdot\pv\,\kv\cdot\rv\right]
 +\rho_{L}(R-P)\left[\po\ro+m^2+\pv\cdot\rv\right] \right\},
\eea
where $p^0=\pm\om_\pv$. 
It is now straightforward to see that the solution of the previous integral 
equation is precisely given by the Ward identity, Eq.~(\ref{eqWI}), 
since after substituting $\mathfrak{D}(r)=i\tw_{\rv}$ inside the integral 
we obtain exactly the expression for the thermal width Eq.~(\ref{eqthermalwidth}). 
We find therefore that the diagrammatic formulation is consistent with the Ward identity.

We note here that the analysis of the Ward identity is easier in the large 
$N_f$ limit than in the weakly coupled limit in the leading logarithmic 
approximation \cite{Aarts:2002tn}. The reason is that the soft-fermion 
contribution in the leading-log approximation, which leads to an 
additional diagram in the integral equation (see Ref.\ \cite{Aarts:2002tn} 
for further details), is subleading in the large $N_f$ limit and does not 
have to be considered.


\section{Color factors}
\label{color}

The results for large $N_f$ QCD can be obtained from the previous 
analysis, provided that the appropriate color factors and electric 
charges 
of the quarks are inserted. The relevant group factors are
 \be
 C_{F}=\frac{N_{c}^{2}-1}{2N_{c}}, 
 \;\;\;\;\;\;\;\;\;\;\;\;
 T_{F}=\frac{1}{2},
 \;\;\;\;\;\;\;\;\;\;\;\;
 d_F=N_c.
\ee
In terms of the effective coupling \cite{Moore:2001fg}
\be
 g^2_{\rm eff} = g^2 T_F,
\ee
where $g$ is the QCD coupling rescaled with $1/\sqrt{N_f}$, so that it 
remains finite in the large $N_f$ limit (cf.\ Eq.\ (\ref{eqcov})), the QCD 
expressions follow from the abelian results with the replacements
\bea
 &&\hm\mbox{quark self energy/thermal width:} 
 \hspace*{1.2cm}  
 e^{2} \longrightarrow g^{2}_{\rm eff}\frac{C_F}{T_F} \\
 &&\hm\mbox{gluon self energy:} 
 \hspace*{4cm}  
 e^{2} \longrightarrow g^{2}_{\rm eff}  \hspace*{1cm}\\
 &&\hm\mbox{``line" piece in the integral equation:} 
 \hspace*{0.6cm}  
 e^{2} \longrightarrow g^{2}_{\rm eff}\frac{C_F}{T_F}   \\
 &&\hm\mbox{``box" piece in the integral equation:} 
 \hspace*{0.6cm}  
 e^{4} \longrightarrow g^{4}_{\rm eff}\frac{C_F}{T_F} 
\eea
If we now define $\chi$ as (compare with Eqs.\ 
(\ref{eqchishear},\ref{eqchiconduct}))
\bea
 \chi(p) = &&\hm \frac{p^2}{\om_\pv} \frac{\cD(p)}{\Gamma_\pv}\frac{C_F}{T_F} 
 \;\;\;\;\;\;\mbox{(shear viscosity)}, \\
 \chi(p) = &&\hm \frac{p}{\om_\pv} \frac{\cD(p)}{\Gamma_\pv}\frac{C_F}{T_F} 
 \;\;\;\;\;\;\mbox{(electrical conductivity)},
\eea
the integral equations are exactly the same as in QED, with no explicit color factors. 
The transport coefficients are then given by
\be
 \eta=-\frac{d_F T_F}{C_F} \frac{2N_f}{15\pi^2}\int_0^\infty 
	dp\,\frac{p^4}{\om_\pv}n_F'(\om_\pv)\chi(p), 
\ee
and
\be
 \sigma=-\frac{d_F T_F}{C_F} \frac{2q_f^2N_f}{3\pi^2}\int_0^\infty 
	dp\,\frac{p^3}{\om_\pv}n_F'(\om_\pv)\chi(p),
\ee
where $q_f$ is the electric charge of the quarks.


\section{Variational solution}
\label{solution}

In order to obtain the shear viscosity and electrical conductivity for 
general values of the effective coupling constant and mass parameter, we 
solve the problem of extremizing the functionals $Q$ variationally.
Following Refs.\ \cite{Arnold:2000dr,Moore:2001fg,Aarts:2004sd}, 
we expand $\chi(p)$ in a finite set of suitably chosen basis functions
$\phi^{(m)}(p)$:
\be
 \chi(p) = N_f \sum_{m=1}^{N_{\rm var}} a_m \phi^{(m)}(p),
\ee
where we factored out an explicit factor of $N_f$, so that the integrals
below are $N_f$ independent. With this Ansatz, the functional $Q$ reads
\be
 Q[\chi] = N_f \sum_m a_m \left[ {\cal S}_m + \half \sum_{n} a_n
 \left( - {\cal F}_{mn} + {\cal H}_{mn} \right)\right],
\ee
with
\bea \nn
 {\cal S}_m =&&\hm  \int_0^{\infty} dp\,\,
  {\cal S}(p) \phi^{(m)}(p) , \\
 \label{eqmn}
 {\cal F}_{mn}  =&&\hm N_f \int_0^{\infty} dp\,\,
 {\cal F}(p)   \phi^{(m)}(p) \phi^{(n)}(p), \\
 \nn
 {\cal H}_{mn} =&&\hm N_f \int_0^{\infty} dp\,\int_0^{\infty} dr\,
 {\cal H}(p,r)  \phi^{(m)}(p) \phi^{(n)}(r).
\eea
Extremizing the functional with respect to the variational parameters
$a_m$ gives the solution
\be
 a_m = \sum_n ( {\cal F} - {\cal H})^{-1}_{mn} {\cal S}_n,
\ee
so that the shear viscosity and electrical conductivity are given by
\be
 \eta = -\frac{d_F T_F}{C_F} \frac{2N_f^2}{15\pi^2}\sum_m {\cal S}_m a_m, 
\;\;\;\;\;\;\;\;
 \sigma = 
 -\frac{d_F T_F}{C_F} \frac{2q_f^2 N_f}{3\pi^2}\sum_m {\cal S}_m a_m,
\ee
with the corresponding values of ${\cal S}_m$ and $a_m$ for each transport 
coefficient. ${\cal S}_m$ is a 1-dimensional integral, ${\cal F}_{mn}$ and
${\cal H}_{mn}^{\rm line}$ are 3-dimensional integrals and for the shear 
viscosity ${\cal H}_{mn}^{\rm box}$ is a 4-dimensional integral. These 
integrals are done with numerical quadrature. The uncertainty due to the 
numerical integration is estimated to be on the percent level or less.

We now discuss the choice of trial functions. We work with the set
\bea
 \phi^{(m)}(p)=&&\hm(p/T)^2\tilde\phi^{(m)}(x)
 \;\;\;\;\;\;\mbox{for}\;\eta, 
 \\
 \phi^{(m)}(p)=&&\hm p/T^2\tilde\phi^{(m)}(x)
 \;\;\;\;\;\;\;\;\;\mbox{for}\;\sigma.
\eea
where $x=p/\sqrt{T^2+mT}$. The prefactors of $p$ are motivated by an 
approximate analytical solution of the integral equation for massless 
fermions in the leading log approximation for asymptotically large 
momentum (see Ref.~\cite{Aarts:2002it} for a similar analysis for $N_{f}=1$). 
The dimensionless argument $x$ is chosen such that it represents the 
magnitude of the typical momentum, which is $p\sim T$ for small 
fermion masses and $p\sim \sqrt{mT}$ for larger masses. 
For the dimensionless functions $\tilde\phi^{(m)}(x)$ we use 
\be
 \tilde\phi^{(m)}(x) = \frac{1}{(1+x)^{m-1}} \sum_{k=0}^{m-1} (-x)^k
\ee
for not too large fermion masses, $m/T\lesssim 5$. This set was 
also used in Ref.~\cite{Aarts:2004sd} for the calculation of the 
shear viscosity in the $O(N)$ model. For larger mass, however, we find 
that this choice of trial functions leads to slow convergence in the 
number of trial functions. In this case we use Laguerre polynomials, 
$\tilde\phi^{(m)}(x) = L_{m-1}(x)$. The results shown in the figures are 
obtained with a set of dimension 4, and the effect of using this truncated 
basis set is smaller than the width of the lines.

\begin{figure}[p]
\centerline{\epsfig{figure=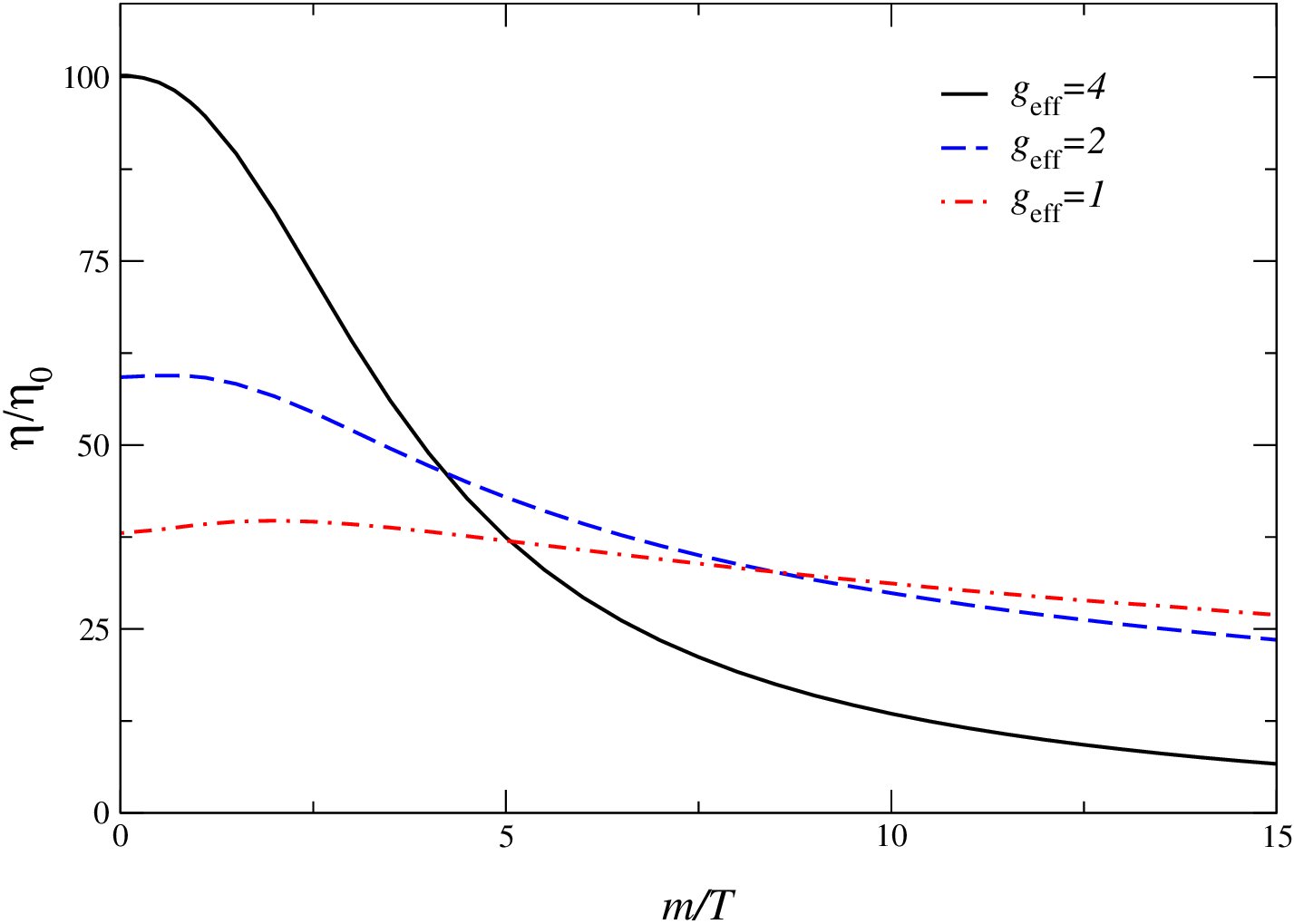,height=7.7cm}}
\caption{Shear viscosity vs.\ the fermion mass $m$ for various values of 
        $g_{\rm eff}$.}
\label{figshear1}
\vspace*{1.35cm}
\centerline{\epsfig{figure=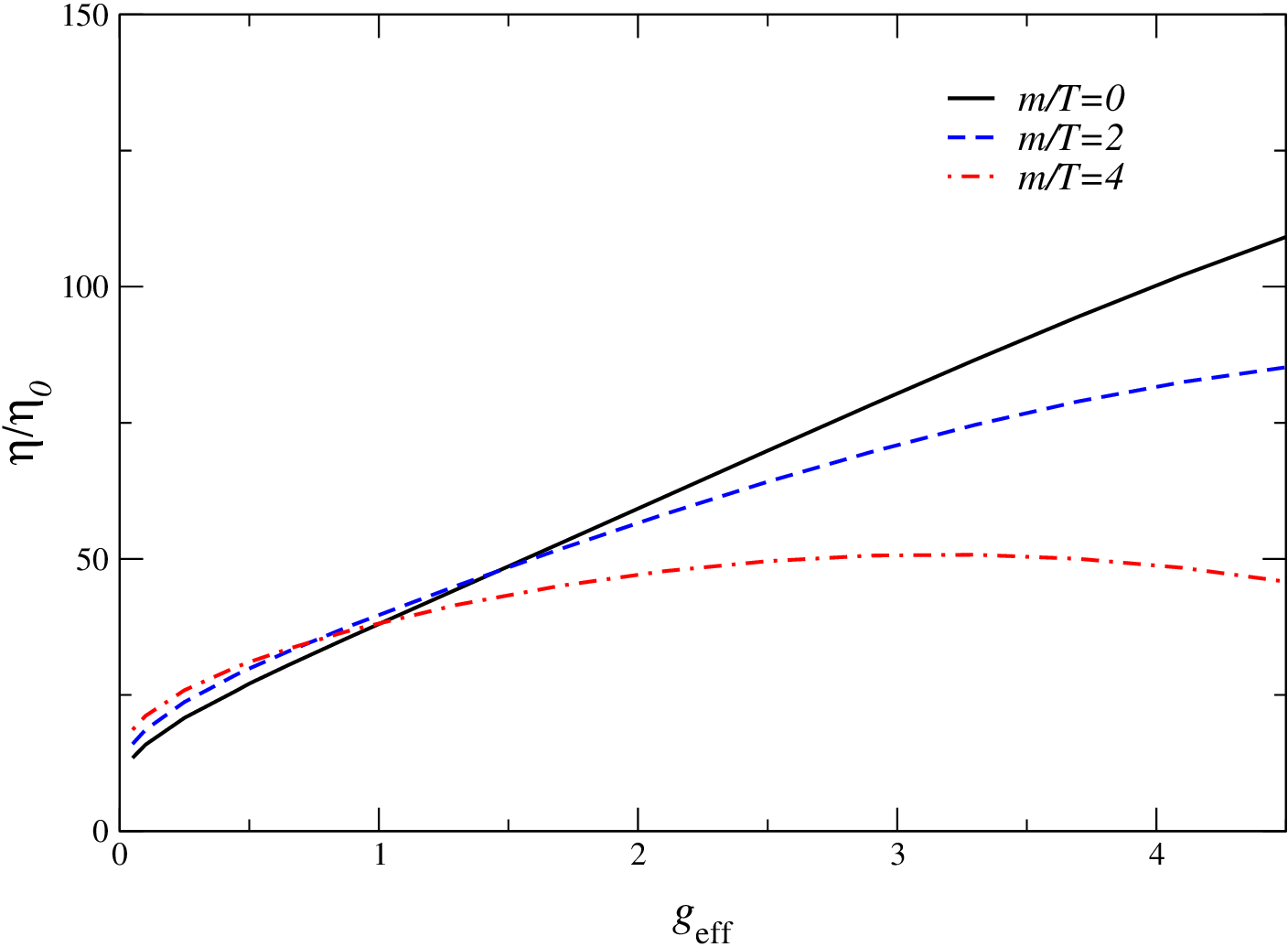,height=7.7cm}}
\caption{Shear viscosity vs.\ the effective coupling constant $g_{\rm 
         eff}$ for various values of the fermion mass $m$.}
\label{figshear2}
\end{figure}

\begin{figure}[p]
\centerline{\epsfysize=7.7cm\epsfbox{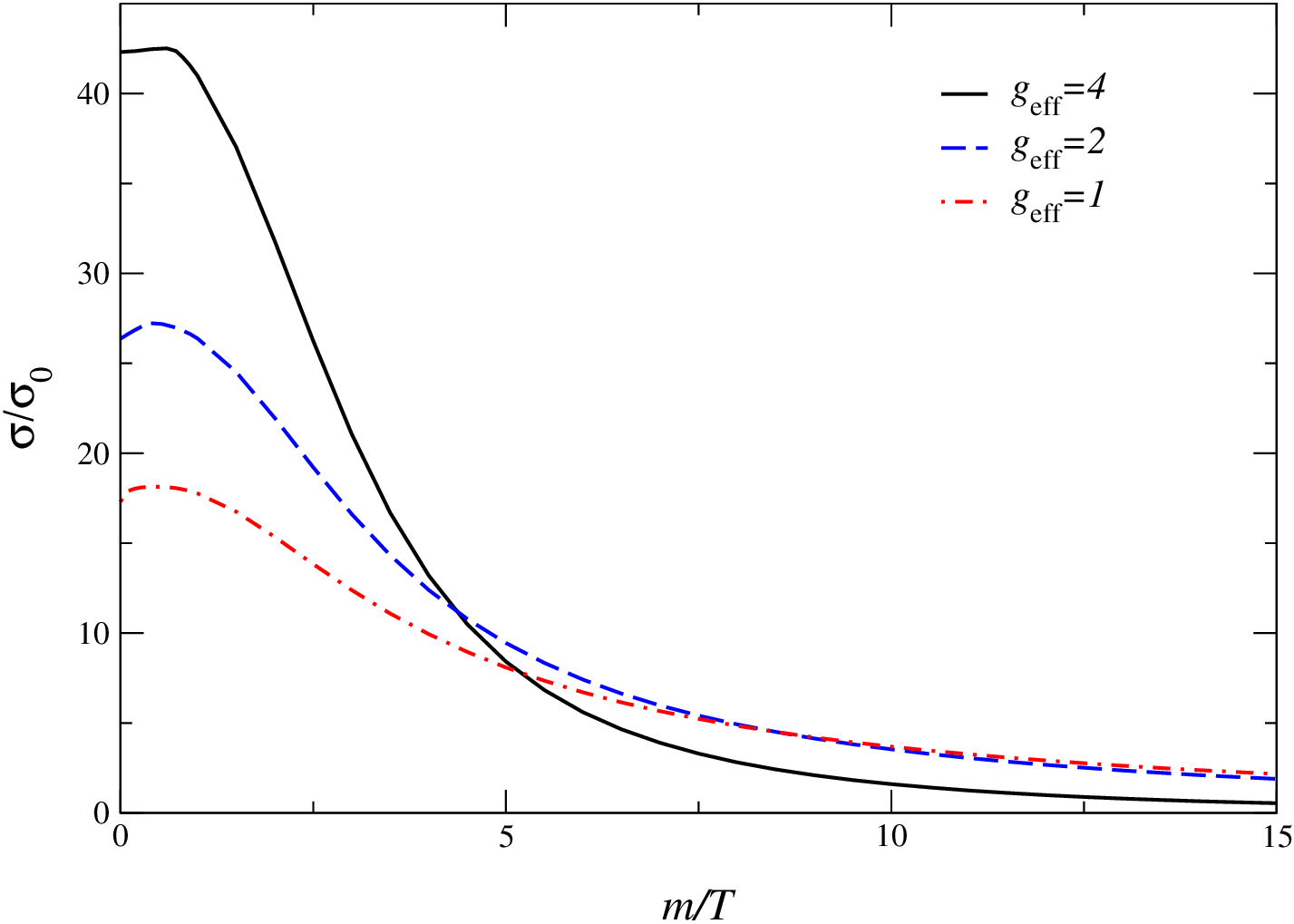}}
\caption{Electrical conductivity vs.\ the fermion mass $m$ for various 
         values of $g_{\rm eff}$.}
\label{figcond1}
\vspace*{1.35cm}
\centerline{\epsfysize=7.7cm\epsfbox{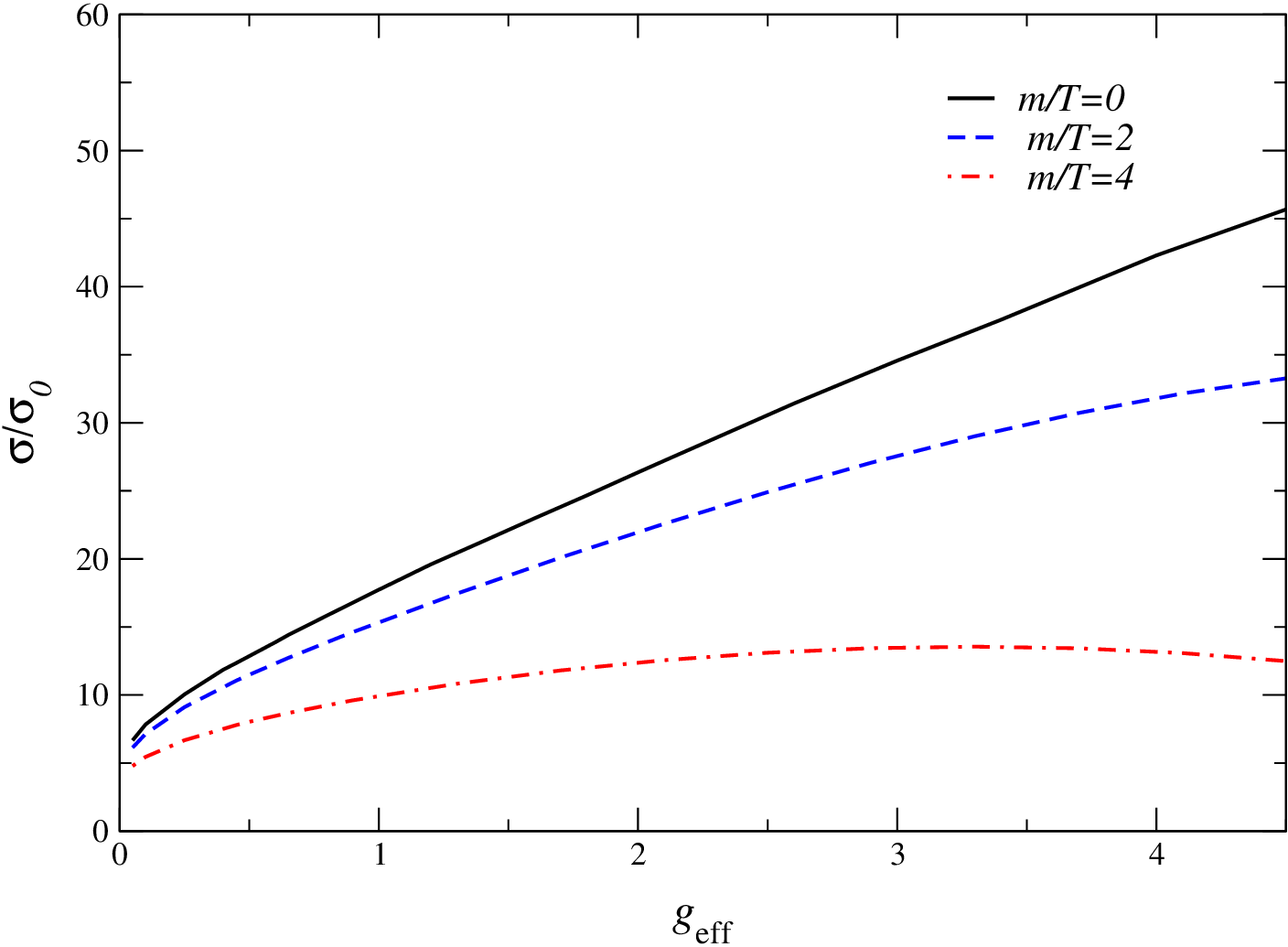}}
\caption{Electrical conductivity vs.\ the effective coupling constant 
         $g_{\rm eff}$ for various values of the fermion mass $m$.}
\label{figcond2}
\end{figure}

The dependence of the shear viscosity on the fermion mass and the 
effective gauge coupling is shown in Figs.~\ref{figshear1} and 
\ref{figshear2}, while the results for the electrical conductivity are 
shown in Figs.~\ref{figcond1} and \ref{figcond2}. The results are 
normalized with
\be
 \eta_0 = \frac{d_FT_F}{C_F}\, \frac{N_f^2}{g_{\rm eff}^4}\, T^3,   \hs{2cm}
 \sigma_0 = \frac{d_FT_F}{C_F}\,\frac{q_f^2N_f^2}{g_{\rm eff}^4}\, T.
\ee
 For QED, $g_{\rm eff}=e$ and $d_F=T_F=C_F=1$, while for QCD, $g_{\rm 
eff}^2=g^2T_F$. We remind that both $e$ and $g$ remain finite in the large 
$N_{f}$ limit and that for the purpose of presentation we have chosen 
$\mu=\mu_{DR}=\pi e^{-\gamma_E}T$.

We notice that the general behavior of these transport coefficients is to 
decrease with increasing mass except for small values of the coupling 
constant, where a slight increase for small masses is observed. 
After rescaling with $\eta_0$ resp.\ $\sigma_0$, the 
remaining dependence on the effective coupling constant 
is quite strong, much stronger than the 
subleading dependence on the quartic coupling in the case of the shear 
viscosity in the $O(N)$ model \cite{Aarts:2004sd}. In the limit of 
vanishing fermionic mass, our results agree with those obtained in 
Ref.~\cite{Moore:2001fg} using kinetic theory. In the opposite limit of 
very large mass, we provide in Appendix \ref{appendixB} some parametric 
estimates of these transport coefficients in the leading logarithmic 
approximation, see Eq.~(\ref{eqLL}), which corroborate the behavior shown 
in the plots. The difference between the mass dependence of the shear 
viscosity in the $O(N)$ model \cite{Aarts:2004sd} and in the gauge theory 
is due to the different mass and momentum dependence of the 
(transport) cross section.

In our complete leading order calculation in the large $N_f$ limit, the 
presence of the Landau pole means that, for given fermion mass, there is 
an upper bound on the possible values of the coupling constant, arising 
from the requirement that all physical scales lie well below the Landau 
scale $\Lambda_L$. For large masses $m/T\gg 1$, the typical momentum scale 
is not set by the temperature, but instead by $p\sim \sqrt{mT} \gg T$. 
Thus, the upper bound on the coupling constant decreases when the mass is 
increased. This implies that going to asymptotically large mass requires a 
restriction to the weak coupling limit.


\section{Conclusions}
\label{conclusions}

We have presented a diagrammatic calculation of the shear viscosity and 
the electrical conductivity at leading order in the large $N_{f}$ 
expansion of QED and QCD for massive fermions. 

The 2PI effective action at next-to-leading order provides in a 
straightforward manner appropriate integral equations which sum all the 
required diagrams to obtain these transport coefficients at leading order. 
We proved that these equations are gauge fixing independent and consistent 
with the Ward identity. This explicitly shows that in a fully 
self-consistent calculation of these transport coefficients at leading 
order in the 2PI framework, potential non gauge invariant contributions 
would be suppressed by powers of $N_{f}$. This suggests that in 
self-consistent applications of the 2PI effective action in gauge 
theories, e.g.\ as in far-from-equilibrium applications, potential 
problems related to gauge invariance and Ward identities would be small 
for sufficiently large $N_f$.

Our results show a nontrivial dependence of the shear viscosity and 
electrical conductivity on the mass of the fermions and the effective 
gauge coupling. We found that for small values of the coupling constant 
they increase slightly with increasing mass. For larger values of the 
fermion 
mass, both the shear viscosity and electrical conductivity decrease. It 
would therefore be interesting to extend current calculations of transport 
coefficients to include different massive fermion flavors. We also found 
that after taking out the expected $1/\alpha^2$ dependence, a strong 
dependence on the gauge coupling remains. When nonperturbative results 
obtained from lattice QCD simulations 
\cite{Gupta:2003zh} are compared with 
perturbative ultrarelativistic expressions, these findings should be kept 
in mind.


\vspace*{0.5cm}
\noindent
{\bf Acknowledgments.} Discussions with G.\ Moore and L.\ Yaffe on the 
large mass dependence are gratefully acknowledged. 
J.M.M.R.\ thanks the Physics Department in Swansea for its hospitality during 
the completion of this work.
This research was supported in part by the 
U.~S.\ Department of Energy under Contract No.\ DE-FG02-01ER41190 and  
No.\ DE-FG02-91-ER4069 and in part by the Spanish Science Ministry 
(Grant FPA 2002-02037) and the University of the Basque Country 
(Grant UPV00172.310-14497/2002). G.A.\ is supported by a PPARC Advanced 
Fellowship.


\appendix 


\section{2PI effective action}
\label{appendixA}

In this appendix we summarize some useful exact relations derived from the 
2PI effective action. We consider only bilocal sources, such that the path 
integral is
\be
Z[K,H] = e^{iW[K,H]} = \int {\cal D}\bar\psi {\cal D}\psi {\cal D}\phi\, 
e^{i\left(S[\bar\psi,\psi,\phi] 
+\frac{1}{2}\phi^i K_{ij}\phi^j
+H_{ba}\psi^a \bar\psi^b \right)}.
\ee
 We denote Bose fields collectively with $\phi^i$ and fermion fields with
$\psi^a$. Indices indicate space-time as well as internal indices, and
integration and summation over repeated indices is understood, e.g.,
 \be 
H_{ba}\psi^a \bar\psi^b = 
\int_{xy} H_{\beta\alpha}(y,x)\psi^\alpha(x) \bar\psi^\beta(y) 
= \int_{xy} \tr\, H(y,x)\psi(x)\bar\psi(y).
\ee
The 2PI effective action follows from the Legendre transform
\be
\Gamma[G,S] = W[K,H] - \frac{1}{2}G^{ij}K_{ij} - H_{ba}S^{ab},
\ee
where
\bea
\label{eqWK1}
\frac{\delta W}{\delta K_{ij}} = &&\hm \half G^{ij} = 
\half\bra T_{\cal C}\phi^i\phi^j\ket, 
\;\;\;\;\;\;\;\;\;\;\;\;
\frac{\delta \Gamma}{\delta G^{ij}} = 
-\frac{1}{2}K_{ij}, \\ 
\label{eqWK2}
\frac{\delta W}{\delta H_{ba}} = &&\hm S^{ab} = 
\bra T_{\cal C}\psi^a\bar\psi^b\ket, 
\;\;\;\;\;\;\;\;\;\;\;\;\;\;\;\;
\frac{\delta \Gamma}{\delta S^{ba}} = 
-H_{ab}.
\eea
The effective action is written as
\be
\Gamma[G,S] = 
\frac{i}{2}\Tr \ln G^{-1} + \frac{i}{2}\Tr\, G_0^{-1}(G-G_0)  
-i\Tr \ln S^{-1} -i\Tr\, S_0^{-1}(S-S_0) 
+\Gamma_2[G,S],
\ee
where 
\be
iG_{0\,ij}^{-1} = \frac{\delta^2 S}{\delta\phi^i\delta\phi^j} 
\Big|_{\phi=\psi=\bar\psi=0}, 
\;\;\;\;\;\;\;\;\;\;\;\;
iS_{0\,ab}^{-1} = \frac{\delta^2 S}{\delta\psi^a\delta\bar\psi^b}
\Big|_{\phi=\psi=\bar\psi=0}.
\ee
Varying this effective action once yields
\be
\frac{\delta \Gamma}{\delta G^{ij}} =
 -\frac{i}{2}\left( G^{-1}_{ij} - G_{0\,ij}^{-1} + 
\Pi_{ij} \right), 
\;\;\;\;\;\;\;\;\;\;\;\;
\frac{\delta \Gamma}{\delta S^{ba}} = 
i\left( S^{-1}_{ab} - S_{0\,ab}^{-1} + 
\Sigma_{ab} \right),
\ee
with
\be
\Pi_{ij} = 
2i\frac{\delta\Gamma_2}{\delta G^{ij}}, 
\;\;\;\;\;\;\;\;\;\;\;\;
\Sigma_{ab} = -i\frac{\delta\Gamma_2}{\delta S^{ba}}.
\ee
Varying one more time results in the relations
\bea
\nn
\frac{\delta^2 \Gamma}{\delta G^{ij} \delta G^{kl} }
=&&\hm \frac{i}{4}\left( 
G^{-1}_{ik}G^{-1}_{jl} + G^{-1}_{il}G^{-1}_{jk} 
-\Lambda_{ij;kl}\right), \\
\frac{\delta^2 \Gamma}{\delta S^{ba} \delta S^{dc} }
=&&\hm i\left( 
-S^{-1}_{ad}S^{-1}_{cb} -\Lambda_{ab;cd}\right),  \\
\nn
\frac{\delta^2 \Gamma}{\delta G^{ij} \delta S^{ba} }
=&&\hm \frac{i}{2}\Lambda_{ij;ab},
\eea
where we defined
\be
\Lambda_{ij;kl} = 4i \frac{\delta^2 \Gamma_2}{\delta G^{ij} \delta 
G^{kl}},
\;\;\;\;\;\;\;\;
\Lambda_{ij;ab} = -2i \frac{\delta^2 \Gamma_2}{\delta G^{ij} \delta 
S^{ba}},
\;\;\;\;\;\;\;\;
\Lambda_{ab;cd} = i \frac{\delta^2 \Gamma_2}{\delta S^{ba} \delta S^{dc}}.
\ee
{}From Eqs.\ (\ref{eqWK1}, \ref{eqWK2}) we find 
\bea
\nn
\frac{\delta^2 W}{\delta K_{ij}\delta K_{kl} }
=&&\hm \frac{i}{4}\left( G_c^{ij;kl} + G^{ik}G^{jl} + 
G^{il}G^{jk} \right), \\
\frac{\delta^2 W}{\delta H_{ba}\delta H_{dc} }
 =&&\hm  i\left( G_c^{ab;cd} - S^{cb}S^{ad} \right), \\
\nn
\frac{\delta^2 W}{\delta K_{ij}\delta H_{ba} } =&&\hm \frac{i}{2} 
G_c^{ij;ab}, 
\eea
where $G_c^{ij;kl}$ etc.\ are the usual connected 4-point functions. 
It is convenient to use vertex functions defined by truncating legs
\bea
\nn
G_c^{ij;kl;} =&&\hm G^{ii'}G^{jj'}G^{kk'}G^{ll'} 
\Gamma^{(4)}_{i'j';k'l';},  \\
G_c^{ij;ab} =&&\hm G^{ii'}G^{jj'}S^{aa'}S^{b'b} 
\Gamma^{(4)}_{i'j';a'b'}, \\
\nn
G_c^{ab;cd;} =&&\hm S^{aa'}S^{b'b}S^{cc'}S^{d'd} 
\Gamma^{(4)}_{a'b';c'd'}.
\eea
{}From the following identities
\be
0 = \frac{\delta K_{ij}}{\delta H_{ba}}, 
\;\;\;\;\;\;\;\;
\delta_a^d\delta_b^c =  \frac{\delta H_{ab}}{\delta H_{dc}}, 
\;\;\;\;\;\;\;\;
0 = \frac{\delta H_{ab}}{\delta K_{ij}}, 
\;\;\;\;\;\;\;\;
\half\left(\delta_i^k\delta_j^l + \delta_i^l\delta_j^k\right) =
\frac{\delta K_{ij}}{\delta K_{kl}},
\ee
we arrive at four coupled integral equations for the 4-point vertex functions:
\bea  \label{eqfourex}
\nn\Gamma^{(4)}_{ij;ab} =&&\hm \Lambda_{ij;ab} 
+ \half \Lambda_{ij;kl} G^{kk'}G^{ll'} \Gamma^{(4)}_{k'l';ab}
- \Lambda_{ij;dc} S^{cc'}S^{d'd} \Gamma^{(4)}_{c'd';ab}, \\
\nn\Gamma^{(4)}_{ab;cd} =&&\hm \Lambda_{ab;cd} 
+\half \Lambda_{ab;ij} G^{ii'}G^{jj'} \Gamma^{(4)}_{i'j';cd}
- \Lambda_{ab;fe} S^{ee'}S^{f'f} \Gamma^{(4)}_{e'f';cd}, \\
\nn\Gamma^{(4)}_{ab;ij} =&&\hm \Lambda_{ab;ij} 
+ \half \Lambda_{ab;kl;} G^{kk'}G^{ll'} \Gamma^{(4)}_{k'l';ij}
- \Lambda_{ab;dc} S^{cc'}S^{d'd} \Gamma^{(4)}_{c'd';ij}, \\
\Gamma^{(4)}_{ij;kl} =&&\hm \Lambda_{ij;kl} 
+ \half \Lambda_{ij;mn} G^{mm'}G^{nn'} \Gamma^{(4)}_{m'n';kl}
- \Lambda_{ij;dc} S^{cc'}S^{d'd} \Gamma^{(4)}_{c'd';ij}.
\eea
In the main text we employ these equations using the $1/N_f$ expansion of 
the 2PI effective action at NLO.


\section{Parametric estimates}
\label{appendixB}

In this appendix we discuss parametric estimates in the leading 
logarithmic approximation, in the zero and large fermion mass limit, using 
a standard kinetic theory discussion (see e.g.\ \cite{LL}).

It follows from the hydrodynamical definitions \cite{Kadanoff} that 
the shear viscosity and the electrical conductivity are related to a 
diffusion coefficient as
\be
D  = \frac{\eta}{\bra {\cal E} + {\cal P}\ket},
\;\;\;\;\;\;\;\;\;\;\;\;\;\;\;
D = \frac{\sigma}{\Xi},
\ee
 where $\cal E$ is the energy density, $\cal P$ the pressure, and $\Xi$ 
the charge susceptibility. For parametric estimates, this diffusion 
constant can be taken to be the same, since similar processes determine 
the transport of energy momentum and charge in the large $N_f$ limit, 
namely large angle scattering between fermions. The diffusion constant can 
be estimated using a random walk model \cite{Forster,Jeon:if} as $D = 
\ell_{\rm mf} \bar v$, where $\bar v$ is the average speed and 
$\ell_{\rm 
mf} \sim 1/ \bar n \sigma_{\rm tr}$ the mean free path, with $\bar n$ the 
mean density and $\sigma_{\rm tr}$ the transport cross section \cite{LL}
\be
 \sigma_{\rm tr} = \int d\Omega \left( 1-\cos\theta\right)
 \frac{d\sigma}{d\Omega}.
\ee
If we keep only the most divergent term in the differential cross section, 
\be
\frac{d\sigma}{d\Omega} \sim 
 \frac{\alpha^2}{N_f^2} \frac{1}{s} \frac{1}{\theta^4} 
\;\;\;\;\;\;\;\; 
(p\gg m), 
\;\;\;\;\;\;\;\;\;\;\;\;
\frac{d\sigma}{d\Omega} \sim 
 \frac{\alpha^2}{N_f^2} \frac{m^2}{p^4} \frac{1}{\theta^4} 
\;\;\;\;\;\;\;\; 
 (p\ll m),
\ee
where $\alpha=e^2/4\pi$, and we recall that the gauge coupling $e^2$ is 
rescaled with $1/N_f$, the transport cross section reads 
\be
 \label{eqtr}
 \sigma_{\rm tr} \sim
 \frac{\alpha^2}{N_f^2 T^2}\int d\theta \frac{1}{\theta} =
 \frac{\alpha^2}{N_f^2 T^2}\ln \theta_{\min}^{-1}.
\ee
Here we used that in the relativistic case the typical momentum $p \sim 
\sqrt{s} \sim T$ while in the nonrelativistic case  $p \sim  m\bar v \sim 
\sqrt{Tm}$. Eq.\ (\ref{eqtr}) holds for both $p\gg m$ and $p\ll m$. 
The divergence at small angles is, to leading log accuracy, cut off by 
Debye screening \cite{LL}.  

For light fermions ($T\gg m$) we use that
\be
\bra {\cal E} \ket \sim \bra {\cal P} \ket \sim N_fT^4,
\;\;\;\;\;\;\;\;\;\; 
\bar  n \sim N_fT^3,
\;\;\;\;\;\;\;\;\;\; 
\Xi \sim N_fq_f^2T^2,
\;\;\;\;\;\;\;\;\;\; 
\bar v \sim 1,
\ee
as well as that $\theta_{\rm min} \sim q/p \sim m_D/T \sim e$, where $q$ 
is the exchanged momentum. This leads to the well-known parametric estimates
\cite{Baym:1990uj,Arnold:2000dr}	
\be
\eta \sim N_f^2\frac{T^3}{\alpha^2\ln 1/\alpha}, 
\;\;\;\;\;\;\;\;\;\;\;\;\;\;\;\;\;\;\;\;\;\;\;\;\;\;\;\;
\sigma \sim q_f^2N_f^2\frac{T}{\alpha^2\ln 1/\alpha}.
\ee

In the case of heavy fermions ($m/T \gg 1$) in the regime where scattering 
can be treated classically ($\alpha^2m/T \gg 1$) \cite{LL}, the exchanged 
momentum $q$ can be estimated as the product of the force $\alpha/r_D^2$ 
and the transit time $r_D/\bar v$ for a passage at impact parameter $r_D$, the 
typical Debye distance \cite{LL}. This yields $q\sim \alpha/(r_D\bar v)$. 
We find therefore that $\theta_{\rm min} \sim q/p \sim \alpha/(r_DT)$. The 
inverse Debye mass $r_D=1/m_D$ is determined from $m_D^2 = 
-\Pi^{00}(p^0=0,\pv \to 0)$. For large $m\gg T$ we find an exponentially 
small Debye mass,
 \be
 m_D^2 \sim \alpha m^2 \left(\frac{T}{m}\right)^{\frac{1}{2}} e^{-m/T} 
\left[1+\cO\left(T/m\right)\right],
\ee
such that
\be
\log \theta_{\rm min}^{-1} \sim \frac{m}{2T} +\frac{3}{4} \log 
\frac{T}{\alpha^2m} + \cO(T/m).
\ee
Combining this with
\be
\bra {\cal E} \ket \sim m\bar n,
\;\;\;\;\;\;\;\;\;\; 
\bra {\cal P} \ket \sim T\bar n,
\;\;\;\;\;\;\;\;\;\; 
\Xi \sim q_f^2\bar n/T,
\;\;\;\;\;\;\;\;\;\; 
\bar v \sim \sqrt{T/m},
\ee
yields
\be
\eta \sim 
N_f^2\frac{T^3}{\alpha^2}\left(\frac{T}{m}\right)^{\frac{1}{2}},
\;\;\;\;\;\;\;\;\;\;\;\;\;\;\;\;\;\;\;\;
\sigma \sim q_f^2N_f^2 \frac{T}{\alpha^2}
\left(\frac{T}{m}\right)^{\frac{3}{2}}.
\label{eqLL}
\ee
As indicated above, these expressions are valid in the leading log 
approximation $\log T/\alpha m_D \gg 1$, the large mass limit $m/T \gg 1$, 
as well as the classical scattering limit $\alpha^2m/T \gg 1$.



\begin{thebibliography}{10}


\bibitem{Arnold:2000dr}
P.~Arnold, G.~D.~Moore and L.~G.~Yaffe,
JHEP {\bf 0011}, 001 (2000)
[hep-ph/0010177];
{\em ibid.} {\bf 0305} (2003) 051
[hep-ph/0302165].

\bibitem{Policastro:2001yc}
G.~Policastro, D.~T.~Son and A.~O.~Starinets,
Phys.\ Rev.\ Lett.\  {\bf 87} (2001) 081601
[hep-th/0104066].

\bibitem{Moore:2001fg}
G.~D.~Moore,
JHEP {\bf 0105}, 039 (2001)
[hep-ph/0104121].

\bibitem{ValleBasagoiti:2002ir}
M.~A.~Valle Basagoiti,
Phys.\ Rev.\ D {\bf 66}, 045005 (2002)
[hep-ph/0204334].

\bibitem{Aarts:2002tn}
G.~Aarts and J.~M.~Mart{\'\i}nez Resco,
JHEP {\bf 0211}, 022 (2002)
[hep-ph/0209048].

\bibitem{Boyanovsky:2002te}
  D.~Boyanovsky, H.~J.~de Vega and S.~Y.~Wang,
  Phys.\ Rev.\ D {\bf 67}, 065022 (2003)
  [hep-ph/0212107].

\bibitem{Buchel:2004di}
A.~Buchel, J.~T.~Liu and A.~O.~Starinets,
Nucl.\ Phys.\ B {\bf 707}, 56 (2005)
[hep-th/0406264].

\bibitem{Defu:2005hb}
  H.~Defu,
  hep-ph/0501284.

\bibitem{Peshier:2005pp}
A.~Peshier and W.~Cassing,
hep-ph/0502138.


\bibitem{Moore:2002md}
G.~D.~Moore,
JHEP {\bf 0210}, 055 (2002)
[hep-ph/0209190];
A.~Ipp, G.~D.~Moore and A.~Rebhan,
JHEP {\bf 0301}, 037 (2003)
[hep-ph/0301057];
A.~Ipp and A.~Rebhan,
JHEP {\bf 0306}, 032 (2003)
[hep-ph/0305030].




\bibitem{Berges:2000ur}
J.~Berges and J.~Cox,
Phys.\ Lett.\ B {\bf 517}, 369 (2001)
[hep-ph/0006160];
B.~Mihaila, F.~Cooper and J.~F.~Dawson,
Phys.\ Rev.\ D {\bf 63}, 096003 (2001)
[hep-ph/0006254];
G.~Aarts and J.~Berges,
Phys.\ Rev.\ D {\bf 64}, 105010 (2001)
[hep-ph/0103049];
J.~Berges,
Nucl.\ Phys.\ A {\bf 699}, 847 (2002)
[hep-ph/0105311];
G.~Aarts and J.~Berges,
Phys.\ Rev.\ Lett.\  {\bf 88}, 041603 (2002)
[hep-ph/0107129];
G.~Aarts, D.~Ahrensmeier, R.~Baier, J.~Berges and J.~Serreau,
Phys.\ Rev.\ D {\bf 66}, 045008 (2002)
[hep-ph/0201308];
F.~Cooper, J.~F.~Dawson and B.~Mihaila,
Phys.\ Rev.\ D {\bf 67} (2003) 051901
[hep-ph/0207346];
{\em ibid.} 056003
[hep-ph/0209051];
hep-ph/0502040;
J.~Berges and J.~Serreau,
Phys.\ Rev.\ Lett.\  {\bf 91} (2003) 111601
[hep-ph/0208070];
J.~Berges, S.~Bors\'anyi and J.~Serreau,
Nucl.\ Phys.\ B {\bf 660} (2003) 51
[hep-ph/0212404];
B.~Mihaila,
Phys.\ Rev.\ D {\bf 68}, 036002 (2003)
[hep-ph/0303157];
S.~Juchem, W.~Cassing and C.~Greiner,
Phys.\ Rev.\ D {\bf 69} (2004) 025006
[hep-ph/0307353];
Nucl.\ Phys.\ A {\bf 743}, 92 (2004)
[nucl-th/0401046];
J.~Berges, S.~Bors\'anyi and C.~Wetterich,
Phys.\ Rev.\ Lett.\  {\bf 93}, 142002 (2004)
[hep-ph/0403234];
A.~Arrizabalaga, J.~Smit and A.~Tranberg,
JHEP {\bf 0410}, 017 (2004)
[hep-ph/0409177].



\bibitem{Aarts:2003bk}
G.~Aarts and J.~M.~Mart{\'\i}nez Resco,
Phys.\ Rev.\ D {\bf 68} (2003) 085009
[hep-ph/0303216].


\bibitem{Arrizabalaga:2002hn}
A.~Arrizabalaga and J.~Smit,
Phys.\ Rev.\ D {\bf 66}, 065014 (2002)
[hep-ph/0207044];
E.~Mottola,
in {\em Proceedings of SEWM2002}, Heidelberg, Germany, 2-5 Oct 2002
[hep-ph/0304279];
M.~E.~Carrington, G.~Kunstatter and H.~Zaraket,
hep-ph/0309084;
A.~Peshier,
Phys.\ Rev.\ D {\bf 70}, 034016 (2004)
[hep-ph/0403225];
J.~O.~Andersen and M.~Strickland,
Phys.\ Rev.\ D {\bf 71}, 025011 (2005)
[hep-ph/0406163].

\bibitem{Aarts:2004xs}
G.~Aarts and J.~M.~Mart{\'\i}nez Resco,
in {\em Proceedings of SEWM04}, Helsinki, Finland, 16-19 June 2004
[hep-ph/0409090].

\bibitem{Aarts:2004sd}
G.~Aarts and J.~M.~Mart{\'\i}nez Resco,
JHEP {\bf 0402} (2004) 061
[hep-ph/0402192].




\bibitem{Cornwall:1974vz}
J.~M.~Cornwall, R.~Jackiw and E.~Tomboulis,
Phys.\ Rev.\ D {\bf 10} (1974) 2428;
J.M.\ Luttinger and J.C.\ Ward, Phys.\ Rev.\ {\bf 118} (1960) 
1417; G.\ Baym, Phys.\ Rev.\ {\bf 127} (1962) 1391.


\bibitem{vanHees:2001ik}
H.~van Hees and J.~Knoll,
Phys.\ Rev.\ D {\bf 65}, 025010 (2002)
[hep-ph/0107200];
{\em ibid.} 105005 (2002)
[hep-ph/0111193];
J.~P.~Blaizot, E.~Iancu and U.~Reinosa,
Phys.\ Lett.\ B {\bf 568} (2003) 160
[hep-ph/0301201];
Nucl.\ Phys.\ A {\bf 736}, 149 (2004)
[hep-ph/0312085];
F.~Cooper, B.~Mihaila and J.~F.~Dawson,
Phys.\ Rev.\ D {\bf 70}, 105008 (2004)
[hep-ph/0407119];
J.~Berges, S.~Bors\'anyi, U.~Reinosa and J.~Serreau,
hep-ph/0409123.


\bibitem{smilga}
V.~V.~Lebedev and A.~V.~Smilga,
Physica A {\bf 181} (1992) 187.

\bibitem{Blaizot:1996az}
J.~P.~Blaizot and E.~Iancu,
Phys.\ Rev.\ D {\bf 55}, 973 (1997)
[hep-ph/9607303].



\bibitem{Aarts:2002it}
G.~Aarts and J.~M.~Mart{\'i}nez Resco,
in {\em Proceedings of SEWM2002}, Heidelberg, Germany, 2-5 Oct 2002
[hep-ph/0212268].


\bibitem{Gupta:2003zh}
  S.~Gupta,
  Phys.\ Lett.\ B {\bf 597}, 57 (2004)
  [hep-lat/0301006];
  A.~Nakamura and S.~Sakai,
  Phys.\ Rev.\ Lett.\  {\bf 94}, 072305 (2005)
  [hep-lat/0406009];
G.~Aarts and J.~M.~Mart{\'\i}nez Resco,
JHEP {\bf 0204} (2002) 053
[hep-ph/0203177];
Nucl.\ Phys.\ Proc.\ Suppl.\  {\bf 119} (2003) 505
[hep-lat/0209033].


\bibitem{LL}
E.~M.~Lifshitz and L.~P.~Pitaevskii, 
{\em Physical Kinetics} 
(Pergamon Press, 1981).

\bibitem{Kadanoff}
L.~P.~Kadanoff and P.~C.~Martin, 
Ann.~Phys.\ {\bf 24}, 419 (1963), reprinted as
Ann.~Phys.\ {\bf 281}, 800 (2000).

\bibitem{Forster}
D.~Forster, 
{\em Hydrodynamic Fluctuations, Broken Symmetry and Correlation Functions}
(Addison Wesley, 1989).

\bibitem{Jeon:if}
S.~Jeon,
Phys.\ Rev.\ D {\bf 52}, 3591 (1995)
[hep-ph/9409250].

\bibitem{Baym:1990uj}
G.~Baym, H.~Monien, C.~J.~Pethick and D.~G.~Ravenhall,
Phys.\ Rev.\ Lett.\  {\bf 64} (1990) 1867.








\end{thebibliography}
\end{document}